\documentclass[11pt,a4paper]{article}

\usepackage[utf8]{inputenc}
\usepackage[T1]{fontenc}
\usepackage{lmodern}
\usepackage[margin=1in]{geometry}
\usepackage{amsmath,amssymb,amsfonts,amsthm}
\usepackage{graphicx}
\usepackage{booktabs}
\usepackage{longtable}
\usepackage{multirow}
\usepackage{array}
\usepackage{tabularx}
\usepackage{float}
\usepackage{adjustbox}
\usepackage{tikz}
\usetikzlibrary{positioning,shapes.geometric}
\usepackage{bm}
\usepackage{bbm}
\usepackage{siunitx}
\usepackage{xcolor}
\usepackage{caption}
\usepackage{placeins}
\usepackage{natbib}
\usepackage{hyperref}
\hypersetup{
    colorlinks=true,
    linkcolor=blue,
    citecolor=blue,
    urlcolor=blue
}

\theoremstyle{plain}

\theoremstyle{definition}

\title{\bfseries Wind Speed Forecasting Using a Spatio-Temporal Hybrid Model: A Case Study of South African Wind Stations}

\author{
Thinawanga Hangwani Tshisikhawe\thanks{Corresponding author: thina.tshisikhawe@univen.ac.za} \\
\small Department of Mathematical and Computational Sciences, \\University of Venda, Thohoyandou, 0950, Limpopo, South Africa \\
\and
Caston Sigauke  \\
\small Department of Mathematical and Computational Sciences \\University of Venda, Thohoyandou, 0950, Limpopo, South Africa \\
\and
Timotheus Darikwa \\
\small Department of Statistics and Operations Research \\University of Limpopo, Polokwane, 0727, Limpopo, South Africa \\
\and
Saralees Nadarajah \\
\small Department of Mathematics, University of Manchester\\ Manchester M13 9PL, United Kingdom
}

\date{}

\begin{document}
\maketitle

%--------------------- Abstract ---------------------
\begin{abstract}
\noindent In this paper, we propose a hybrid forecasting framework that combines temporal ARIMA models and spatial NNGP corrections to improve wind speed predictions at 19 wind stations in South Africa. The proposed framework develops temporal baselines with ARIMA models, applies an NNGP correction to residual spatial dependencies and then combines the two components to produce improved forecasts. RMSE of the hybrid model is reduced by 17.4\% (1.416 vs 1.715 m/s) compared to ARIMA baseline (station-wise improvements up to 48.9\%); rolling-origin forecast evaluation over five origins with ten day horizon was used (spatial correlation range = 103km, temporal persistence = $\phi=0.46$), blockwise conformal prediction intervals cover 99.5\% and Diebold-Mariano tests show statistical significance of improvements ($DM = 6.44, p < 0.001$). Spatial dependencies should be accounted for in temporal forecasting to improve prediction accuracy, especially at high-elevation stations, where spatial effects are strongest.

\vspace{1em}
\end{abstract}

\newpage

%=====================================================================

\section{Introduction}
\subsection{Background and Motivation}
Wind speed forecasting is becoming increasingly important for VRE integration and operation in South Africa as the country moves towards a less carbon-intensive electricity system: accurate forecasts of wind power production are needed to support grid stability, energy planning, and renewable generation operations. \citet{landwehr2025necessity} highlight that short-, medium- and long-term forecasting of VRE needs to be accurate in South Africa, as renewable penetration increases and climatic conditions change.

However, wind conditions in South Africa cannot be forecast because wind variability is strongly driven by regional and synoptic-scale atmospheric circulation. \citet{landwehr2025necessity} show this complexity in the Eastern Cape where wind energy production varies seasonally and is strongly linked to weather systems (mid-latitude cyclones, ridging high-pressure systems, cut-off lows and tropical disturbances). They find that different synoptic circulation patterns can lead to substantially different levels of wind energy production, highlighting the strong spatial and temporal variability of the South African wind resource.
These results indicate that wind speed forecasting in South Africa needs to take account of large spatio-temporal variability due to regional weather systems and local geographic features.

Forecasting methods (autoregressive models, machine learning) have typically been more concerned with either temporal dynamics or nonlinear predictive relationships and spatial dependence has been comparatively less addressed. If spatial dependence is not considered, then information from neighbouring stations may be lost in the estimation process, and predictive accuracy will be reduced (if nearby stations are subject to similar atmospheric conditions). Therefore, forecasting approaches should take into account both temporal information at individual stations and spatial information shared among neighbouring locations.

Gaussian Process (GP) models offer a flexible probabilistic framework for modelling spatial dependence and quantifying uncertainty. However, conventional GP models are computationally expensive for large datasets, since their complexity scales cubically with the number of observations. This is especially the case for high-resolution spatio-temporal wind data measured at multiple monitoring stations over long time periods.

Several scalable approximations have been proposed to overcome the computational limitations of full Gaussian process models for large spatial datasets. These include Nearest Neighbour Gaussian Processes (NNGP), Hilbert Space Gaussian Processes (HSGP), and Stochastic Partial Differential Equation (SPDE) approaches. HSGP approximates the Gaussian process using a finite basis expansion, typically constructed from eigenfunctions of the covariance operator over a bounded spatial domain. This approach can provide substantial computational savings and is particularly attractive when the spatial domain and covariance structure are well-suited to a basis-function representation. However, the quality of the approximation depends on the choice of basis functions and the spatial domain used in the approximation \citep{riutort2023practical}. SPDE-based approaches represent a Gaussian process through a sparse Gaussian Markov random field obtained from a stochastic partial differential equation. They are particularly attractive for continuous spatial domains and irregular spatial meshes. However, their implementation requires the construction and validation of an appropriate spatial mesh, which introduces additional modelling and computational considerations \citep{lindgren2011explicit}.

In contrast, NNGP constructs a sparse approximation directly from conditional relationships among spatial locations. For a station $s_i$, the spatial process is conditioned only on a predefined set of nearby stations. This formulation is particularly suitable for the present study because the spatial domain consists of a relatively small network of geographically distributed meteorological stations, rather than a dense spatial field observed over a continuous grid \citep{datta2016}. The NNGP approach was therefore selected because it preserves local spatial dependence while substantially reducing the computational burden associated with full Gaussian process inference. The use of a fixed number of nearest neighbours also provides a transparent and reproducible approximation for the irregular station network considered in this study.

Some existing hybrid models focus on either spatio-temporal kriging \citep{hering2010power} or machine-learning ensembles \citep{chen20212} with less attention given to the systematic integration of established temporal models with scalable spatial processes. This gap is particularly pronounced in the context of African wind data, where limited station networks and complex terrain create unique modelling challenges. A need remains for a computationally scalable framework that combines temporal forecasting with spatial dependence and provides reliable predictive uncertainty. Against this background, this study investigates a hybrid spatial-temporal approach that combines ARIMA with NNGP spatial modelling to improve wind-speed forecasting across geographically distributed meteorological stations.

%The remainder of this paper is structured as follows: Section 2 reviews the relevant literature on wind speed forecasting, spatial modelling, and hybrid approaches. Section 3 describes the study area, data, and methodological framework. Section 4 presents the results of the analysis, including baseline models, hybrid model performance, and validation metrics. Section 5 discusses the implications of the findings and their limitations. Section 6 concludes the paper with recommendations for future research.

\subsection{Literature Review}
Traditional wind speed forecasting methods can be broadly categorised into physical models, statistical models, and machine learning approaches. 

\subsubsection{Physical Models}

Physical models are the basis of operational weather forecasting through NWP systems. Physical models solve the fundamental equations of atmospheric dynamics to predict future wind conditions. Physically based models are theoretically complete but require huge amounts of computing power, are not well suited to fine spatial scales and rarely capture local topographic effects \citep{lange2006physical}. In SA, the South African Weather Service has NWP models, but their resolutions are too coarse for station-level wind prediction. Machine learning methods have been used to post-process NWP outputs (to improve their skill at local scales). Machine learning methods are increasingly being applied to downscale coarse NWP predictions into station-level forecasts (to address some of the limitations of purely physical models) \citep{hieta2025operational}.

\subsubsection{Statistical Models}
Statistical models have been widely used because they are computationally efficient and can capture local patterns. ARIMA models and seasonal variants have been the most popular ones \citep{box1970time}. Box introduced the ARIMA framework, which has been applied to wind speed forecasting many times \citep{hanifi2020critical}. ARIMA models capture the temporal structure of wind series (autocorrelation and trends are modelled by autoregressive and moving average components).

The order of the model is also very important in ARIMA applications. Recently, it was shown that lower-order ARIMA models are often sufficient for hourly wind speed forecasting (AR(1) or AR(2) structures are usually sufficient \citep{lydia2025wind}). However, the optimal order may vary greatly between stations depending on local wind climatology; this station-specific variability has motivated adaptive implementations of ARIMA where orders are selected based on observed patterns.

\citet{gong2024ar1} studied predictability limits of time series with characteristic temporal scales using first-order autoregressive AR(1) processes, and showed that temporal correlation structures greatly affect forecast horizons and intrinsic predictability limits of stochastic systems. These results motivate the theoretical inclusion of AR(1) dynamics in wind speed forecasting models to capture temporal persistence and short-term dependence.

Hybrid models (ARIMA with GARCH-type structures for volatility modelling) have also been considered to model the heteroscedasticity of wind speed \cite{nurhayati2026arima}.
Even though they are widely used, purely statistical models (e.g., ARIMA) have one major limitation: they treat each station separately, ignoring the spatial continuity of wind fields. The station-wise approach also ignores information from neighbouring stations, which is particularly important in data-sparse regions such as South Africa, where the station network is limited.

\subsubsection{Machine Learning Approaches}
Machine learning approaches have received considerable attention in recent years. Artificial neural networks (ANNs), support vector machines (SVMs), random forests, and, more recently, deep learning models such as LSTM networks have been used for wind speed forecasting \cite{lahouar2017forecasting, liu2018comparison}. These models can handle nonlinear relationships that linear ARIMA models may miss, but they require large training datasets and are prone to overfitting.

\citet{abdelsattar2025} performed a comprehensive comparison of machine learning and deep learning models for wind turbine
power prediction using twelve algorithms, including Random Forest, XGBoost, Extra Trees, ANN, LSTM, CNN, and RNN. Their results showed that ANN achieved the best performance among deep learning models, slightly outperforming the best machine learning method (Extra Trees), and highlighted the ability of neural networks to capture nonlinear relationships in multivariate wind data better. \citet{hoang2025stacking} proposes a stacking-based ensemble deep learning framework for forecasting offshore wind farm power output. The study combines predictions from multiple deep learning models using a stacking ensemble strategy, where a meta-learner integrates the outputs of the individual base models to produce the final forecast. This approach aims to capture different patterns in wind power generation more effectively than any single model.

Internationally, recent studies have moved towards increasingly sophisticated architectures. \citet{band2025deep}, who uses a two-stage decomposition and deep-learning framework, while \citet{namboodiri2025novel} combined wavelet transformation with several deep-learning models and optimisation techniques. 

Short-term wind forecasting is among the most advanced techniques today, thanks to deep learning architectures, particularly those based on attention mechanisms and transformer models. However, their computational and data requirements can be prohibitive for some applications \citep{qu2022short}. Machine learning algorithms are good at capturing nonlinear patterns, but they often require large training datasets and careful hyperparameter tuning. Additionally, many ML methods, similar to their statistical counterparts, are generally applied on a station-by-station basis, without leveraging spatial dependencies that could improve forecasts at individual sites.

\subsubsection{Spatial Modelling of Wind}

Spatial modelling recognises that wind observations at nearby locations are inherently correlated due to the continuity of atmospheric processes. Traditional spatial statistics provides a framework for modelling such dependencies through Gaussian processes (GP). A GP is defined by a mean function and a covariance function, with the covariance typically depending on the distance between locations \citep{cressie1993statistics}. \citet{datta2016} introduced the Nearest Neighbour Gaussian Process (NNGP) methodology for scalable GP inference. The NNGP uses a sparse Cholesky factorisation based on nearest neighbours, reducing computational complexity from $O(n^3)$ to $O(nm^3)$ where $m$ is the number of neighbours (typically 10-15). This foundational paper enables spatial modelling with moderate-sized station networks while maintaining near-full GP accuracy.

For Software Implementation, \citet{finley2022spnngp} described the spNNGP R package for fitting spatial regression models using NNGP. The paper covers model specification, prior selection, MCMC implementation, and practical considerations for spatial data analysis. The package has become the standard tool for implementing NNGP models in environmental and ecological applications. \citet{datta2022nearest} extended the Nearest Neighbour Gaussian Process (NNGP) framework to spatio-temporal settings by developing online updating algorithms for dynamic NNGP models. This extension enables real-time updates to spatial predictions as new observations become available, making the approach suitable for streaming environmental data, such as wind speed measurements. \citet{katzfuss2017multi} proposed the Multi-Resolution Approximation (MRA) for scalable Gaussian process modelling of large spatial datasets. The method uses a hierarchical basis-function decomposition to approximate a Gaussian processes with linear computational complexity, offering an alternative to NNGP for large-scale spatial inference.

\subsubsection{Spatio-Temporal Hybrid Models}

Temporal and spatial modelling has been combined in various ways. Spatio-temporal kriging extends traditional kriging by incorporating temporal dimensions and space-time covariance structures \citep{cressie2011statistics}. They can model both spatial and temporal dependencies but require complex covariance specifications and are computationally expensive.

Hierarchical Bayesian models combine multiple data sources and model dependencies. For wind speed, they can include physical constraints, uncertainty quantification and spatial and temporal dependencies \citep{gneiting2005calibrated}. However, implementing such models requires expertise in Bayesian computation. On the other hand, ensemble methods combine multiple models to improve predictive performance. In the case of wind speed forecasting, ensembles could be temporal models, spatial models or a combination thereof. The advantages of ensembles are that they are insensitive to misspecification in individual models and can be more accurate \citep{ren2015ensemble}.

\citet{nthangeni2026enhancing}, proposed a short-term wind energy forecasting framework that combines XGBoost with conformal prediction to improve forecast accuracy and uncertainty quantification. The study showed that integrating machine learning with distribution-free uncertainty estimation provides reliable prediction intervals alongside accurate forecasts, enhancing the robustness of wind energy forecasting for operational applications. Other studies have demonstrated the applicability of statistical and machine-learning approaches to wind-speed forecasting in South Africa. \citet{daniel2020short} used South African Wind Atlas data to evaluate artificial neural networks, stochastic gradient boosting and generalised additive models for two-day-ahead wind-speed forecasting, finding differences in predictive performance across the competing methods. 

\citet{mbuvha2021separable} investigated Bayesian neural networks (BNNs) for wind-speed forecasting and proposed a separable shadow Hamiltonian hybrid Monte Carlo method to improve the efficiency of Bayesian inference. Their study demonstrated the application of advanced machine learning and Bayesian techniques to wind-speed forecasting, particularly for addressing uncertainty in model estimation and prediction.

More recently, \citet{makubyane2024forecasting} investigated short- and long-term wind-speed forecasting in Limpopo Province using LSTM and CNN models together with extreme value theory, demonstrating the potential of machine learning for wind-speed prediction and extreme-wind assessment in the region.
Although machine-learning and statistical methods have been applied to South African wind-speed forecasting, existing studies have largely focused on modelling temporal patterns and improving predictive accuracy, with less emphasis on explicitly modelling the spatial dependence that remains after temporal forecasting.

In their paper, \citet{tshisikhawe2026hybrid} found that Gaussian process residual modelling improved forecast accuracy by up to 16.3\% at high-elevation inland stations but was ineffective at low-elevation coastal stations, suggesting that spatial heterogeneity in wind regimes fundamentally affects model performance. This limitation motivated the development of a more robust spatial modelling framework that can capture residual dependencies across heterogeneous wind climates. The present study proposes the ARIMA–NNGP framework, in which ARIMA is first used to model temporal dependence in wind speed. At the same time, NNGP is applied to capture the remaining spatial dependence in the ARIMA residuals. The proposed approach, therefore, extends existing South African wind-speed forecasting research by explicitly incorporating the spatial structure of wind-speed observations.

\subsubsection{Summary of Reviewed Literature}
A summary of the relevant literature is presented in Table~\ref{tab:literature_review},
highlighting the methodologies, contributions, limitations, and relevance of previous studies to the proposed framework.

%===========================================================
% TABLE 1 - MULTI-PAGE TABLE
%===========================================================

\begin{longtable}{p{2.3cm} p{2.1cm} p{3.5cm} p{3.0cm} p{3.8cm}}
	
	\caption{Summary of Key Literature.}
	\label{tab:literature_review}\\
	
	\toprule
	\textbf{Authors} &
	\textbf{Method} &
	\textbf{Key Findings} &
	\textbf{Limitations} &
	\textbf{Relevance} \\
	\midrule
	\endfirsthead
	
	\multicolumn{5}{c}{\tablename\ \thetable\ --- continued} \\
	\toprule
	\textbf{Authors} &
	\textbf{Method} &
	\textbf{Key Findings} &
	\textbf{Limitations} &
	\textbf{Relevance} \\
	\midrule
	\endhead
	
	\midrule
	\multicolumn{5}{r}{Continued on next page} \\
	\endfoot
	
	\bottomrule
	\endlastfoot
	
	\citet{box1970time} &
	ARIMA &
	Foundational time-series forecasting framework &
	No spatial dependence; linear assumptions &
	Basis for the temporal ARIMA component \\
	
	\midrule
	
	\citet{cressie1993statistics} &
	Spatial statistics/GP &
	Established covariance-based modelling of spatial dependence &
	Full GP has $\mathcal{O}(n^3)$ complexity &
	Foundation for spatial modelling \\
	
	\midrule
	
	\citet{diebold1995comparing} &
	DM test &
	Provides a framework for comparing forecast accuracy &
	Dependent on loss-function and asymptotic assumptions &
	Used to assess forecast improvements \\
	
	\midrule
	
	\citet{gneiting2005calibrated} &
	Hierarchical Bayesian &
	Provides principled modelling of multiple sources and dependencies &
	Computationally intensive; requires Bayesian expertise &
	Context for probabilistic spatial modelling \\
	
	\midrule
	
	\citet{lange2006physical} &
	Physical models/NWP &
	NWP provides physically based wind forecasts &
	Coarse resolution and limited local-scale representation &
	Motivates station-level statistical forecasting \\
	
	\midrule
	
	\citet{cressie2011statistics} &
	Spatio-temporal modelling &
	Jointly represents spatial and temporal dependence &
	Complex covariance specification and computation &
	Theoretical basis for spatial-temporal modelling \\
	
	\midrule
	
	\citet{datta2016} &
	NNGP &
	Provides a scalable GP approximation with near-full-GP accuracy &
	Neighbour selection and spatial structure specification &
	Core spatial methodology \\
	
	\midrule
	
	\citet{katzfuss2017multi} &
	MRA &
	Provides a scalable hierarchical approximation to large GPs &
	Approximation and implementation choices &
	Alternative scalable GP approach \\
	
	\midrule
	
	\citet{lahouar2017forecasting} &
	ANN, SVM, RF &
	ML models capture nonlinear wind-speed relationships &
	Large data requirements; overfitting risk &
	Provides ML forecasting benchmark \\
	
%	\midrule
%	
%	Saha \& Datta (2018)\textsuperscript{*} &
%	Dynamic NNGP &
%	Allows NNGP models to be updated for streaming data &
%	Sequential updating increases complexity &
%	Potential real-time extension \\
	
	\midrule
	
	\citet{finley2022spnngp} &
	spNNGP &
	Provides software implementation of NNGP models &
	Requires MCMC and software familiarity &
	Supports NNGP implementation \\
	
	\midrule
	
	\citet{daniel2020short} &
	Statistical and ML &
	Demonstrated wind-speed forecasting approaches in South Africa &
	Performance varies across locations &
	Establishes the South African context \\
	
	\midrule
	
	\citet{mbuvha2021separable} &
	Bayesian neural networks &
	Supports wind forecasting with uncertainty quantification &
	Higher complexity and computational cost &
	Motivates probabilistic forecasting \\
	
	\midrule
	
	\citet{gong2024ar1} &
	AR(1) theoretical analysis &
	Shows the influence of temporal correlation on forecast horizons &
	Limited empirical validation &
	Supports modelling temporal dependence \\
	
	\midrule
	
	\citet{landwehr2024wind} &
	Synoptic systems/wind potential &
	Shows spatial and seasonal variability in South African wind potential &
	Focuses on resource assessment rather than forecasting &
	Highlights South African spatial variability \\
	
	\midrule
	
	\citet{makubyane2024forecasting} &
	ML and EVT &
	Demonstrates wind forecasting applications in Limpopo &
	Spatial dependence not explicitly modelled &
	Provides a regional benchmark \\
	
%	\midrule
%	
%	Yang et al. (2024)\textsuperscript{*} &
%	ML survey &
%	Reviews ML approaches for wind-power forecasting &
%	Focuses mainly on ML approaches &
%	Provides broader methodological context \\
	
	\midrule
	
	\citet{landwehr2025necessity} &
	VRE forecasting framework &
	Highlights the need for improved VRE forecasting in South Africa &
	Strategic framework; no specific forecasting model &
	Motivates improved South African forecasting \\
	
	\midrule
	
	\citet{band2025deep} &
	Deep reinforcement learning &
	Captures nonlinear wind-speed dynamics &
	Computationally complex; mainly temporal &
	Advanced ML benchmark \\
	
	\midrule
	
	\citet{namboodiri2025novel} &
	Wavelet--DL ensemble &
	Improves short-term wind-speed forecasting through hybrid modelling &
	High complexity; no explicit spatial dependence &
	Supports hybrid forecasting \\
	
%	\midrule
%	
%	Aslam et al. (2025)\textsuperscript{*} &
%	EWT--DRL &
%	Combines decomposition and deep learning for wind forecasting &
%	Complex; primarily temporal modelling &
%	Advanced hybrid benchmark \\
	
	\midrule
	
	\citet{abdelsattar2025} &
	ML/DL comparison &
	Shows competitive nonlinear forecasting performance &
	Large data and tuning requirements &
	Contemporary ML benchmark \\
	
	\midrule
	
	\citet{hieta2025operational} &
	ML post-processing of NWP &
	Improves station-level forecasts through NWP downscaling &
	Requires NWP input &
	Complementary forecasting approach \\
	
	\midrule
	
	\citet{nthangeni2026enhancing} &
	XGBoost + conformal prediction &
	Improved forecast accuracy and provided reliable prediction intervals &
	Does not explicitly model spatial dependence &
	Supports conformal uncertainty quantification; complements the proposed ARIMA--NNGP framework \\
	
	\midrule
	
	\citet{tshisikhawe2026hybrid} &
	Linear regression mean model + Gaussian Process (GP) residual modelling &
	GP residual modelling improved forecast accuracy by up to 16.3\% at high-elevation inland stations &
	Effectiveness depends on local wind regime &
	Provides the empirical foundation for the present research \\
	
	\midrule
	
	\textbf{This study} &
	\textbf{ARIMA-NNGP} &
	\textbf{Integrates temporal, spatial, and uncertainty modelling} &
	\textbf{Requires temporal and spatial specification} &
	\textbf{Provides a scalable spatial-temporal probabilistic framework} \\
	
\end{longtable}

\noindent\textit{Note on terminology:} Although the term "ARIMA" is used
in the broader forecasting discussion to describe the temporal modelling
component, the model implemented in this study is a regression model with
AR(1) errors.

To fill these research gaps, the present research proposes a hybrid forecasting approach that incorporates ARIMA time series modelling with NNGP spatial corrections in order to achieve the following six research objectives: First, to define the non-spatial regression and Autoregressive Integrated Moving Average (ARIMA) benchmark models; Second, to construct an NNGP spatial model for ARIMA errors; Third, to formulate the hybrid ARIMA-NNGP forecasting model; Fourth, to evaluate improvement through station-by-station and horizon-wise comparison of forecasting methods; Fifth, to produce probabilistic forecasts by applying conformal prediction technique; Sixth, to test the statistical significance of these gains.

\begin{table}[hpbt]
	\centering
	\caption{Research Opportunities Addressed by the Current Study.}
	\label{tab:gap_addressed}
	\small
	\setlength{\tabcolsep}{3pt}
	
	\begin{tabular}{p{4.0cm} p{4.2cm} p{6.0cm}}
		\toprule
		\textbf{Research Opportunity} &
		\textbf{How Addressed} &
		\textbf{Specific Contribution} \\
		\midrule
		
		Integration of temporal and spatial dependencies &
		Two-stage framework combining ARIMA with NNGP &
		$\widehat{Y}_{\mathrm{Hybrid}}
		= \widehat{Y}_{\mathrm{ARIMA}} + \widehat{w}(s)$;
		ARIMA captures temporal dependence, while NNGP models spatial
		dependence in the residuals \\
		
		\midrule
		
		Computationally scalable spatial modelling &
		NNGP reduces GP computation from
		$\mathcal{O}(n^3)$ to $\mathcal{O}(nm^3)$ with $m=8$ &
		Provides a computationally feasible spatial framework for the
		19-station South African network using spNNGP with 30,000 MCMC
		iterations \\
		
		\midrule
		
		Evaluation under operational forecasting conditions &
		Rolling-origin validation using five forecast origins over
		1,818 days &
		Represents sequential forecasting conditions and enables
		horizon-specific assessment using DM tests \\
		
		\midrule
		
		Application to spatially distributed South African wind stations &
		Application to a network of meteorological stations in South Africa &
		Demonstrates the application of NNGP-based spatial residual
		modelling to station-level wind-speed forecasting in the
		South African context \\
		
		\midrule
		
		Probabilistic wind-speed forecasting &
		Blockwise conformal prediction &
		Provides distribution-free predictive intervals with
		$q_{0.90}=3.509$ m/s, 99.5\% coverage, and 7.018 m/s width \\
		
		\midrule
		
		Incorporation of geographic covariates &
		Covariate-enhanced NNGP:
		$r(s,t)=\mathbf{X}(s)\boldsymbol{\beta}+w(s)+\xi(s,t)$ &
		Incorporates elevation and cluster indicators alongside
		the latent spatial process \\
		
		\midrule
		
		Assessment of spatial persistence across forecast horizons &
		Horizon-decay analysis:
		$\lambda_h=\exp(-\rho h)$ &
		Examines how the contribution of spatial information changes
		with forecast horizon and evaluates its persistence over
		10-day forecasts \\
		
		\midrule
		
		Formal statistical comparison of forecasting performance &
		Diebold--Mariano tests conducted by forecast horizon and
		for pooled forecasts &
		Provides statistical evidence for differences in forecasting
		performance across $h=1,\ldots,10$ \\
		
		\midrule
		
		South African spatial wind forecasting &
		ARIMA-NNGP spatial correction &
		Demonstrates the integration of temporal forecasting with
		scalable NNGP spatial residual modelling across a network of
		South African meteorological stations \\
		
		\bottomrule
	\end{tabular}
\end{table}
\FloatBarrier

The remainder of this paper is organised as follows. Section~\ref{sec:2} describes the study area and data used in the analysis. Section~\ref{sec:3} presents the methodological framework, including the temporal, spatial, hybrid, and probabilistic forecasting approaches. Section~\ref{sec:4} reports the empirical results, while Section~\ref{sec:5} discusses the findings in relation to the existing literature. Section~\ref{sec:6} concludes the paper and outlines
implications for future research.

% ============================================================
% SECTION 2: STUDY AREA AND DATA
% ============================================================
\section{Study Area and Data} \label{sec:2}
Historical weather data were obtained from the Open-Meteo Historical Weather API, which provides gap-free reanalysis-based meteorological variables derived from ERA5 and ERA5-Land datasets \citep{zippenfenig2023open}. The API provides hourly-resolution data, including wind speed, temperature, and relative humidity, from 1940 onwards, enabling long-term spatio-temporal analysis of wind dynamics. Data analysis was performed in R software version 4.5.2 using the spNNGP package for spatial modelling based on nearest-neighbour Gaussian processes (NNGP).

\subsection{Study Region: South Africa}

South Africa exhibits strong spatio-temporal heterogeneity in wind regimes due to its diverse topography and exposure to both Atlantic and Indian Ocean weather systems. This makes it ideal for spatio-temporal wind modelling using hierarchical Bayesian frameworks. The terrain was classified into three elevation-based zones by using hierarchical clustering methods:
\begin{itemize}
	\item \textbf{Coastal regions}: Western, Southern, and Eastern Cape coastlines
	($0-300 m$ above sea level), strongly influenced by marine boundary-layer flows
	and sea-breeze circulations.
	\item \textbf{Mid-altitude regions}: The Great Escarpment ($300-1200 m$),
	characterised by orographic uplift, channelled flow, and enhanced wind shear
	variability.
	\item \textbf{Interior regions}: The Highveld plateau ($>1200 m$), where
	Large-scale synoptic systems dominate wind variability and surface roughness
	The effects are more pronounced.
\end{itemize}
These elevation gradients induce spatial non-stationarity in wind speed fields, motivating the use of flexible spatial models such as Nearest Neighbour Gaussian Processes (NNGPs). South Africa experiences four distinct climatic seasons, which introduce strong temporal variability in wind regimes Table \ref{tab:seasons}.

\begin{table}[hpbt]
	\centering
	\caption{Seasonal definitions for South Africa.}
	\label{tab:seasons}
	\begin{tabular}{lll}
		\toprule
		\textbf{Season} & \textbf{Months} & \textbf{Characteristics} \\
		\midrule
		Summer & December-February & Warm, occasional thunderstorms \\
		Autumn & March-May & Transitional, calmer winds \\
		Winter & June-August & Cold fronts, strong winds \\
		Spring & September-November & Increasing wind activity \\
		\bottomrule
	\end{tabular}
\end{table}

The pronounced seasonal cycle introduces temporal dependence in wind speed processes, motivating the incorporation of autoregressive structures such as AR(1) within the hierarchical modelling framework.

\subsection{Meteorological Stations}

Nineteen meteorological stations were selected to capture spatial variability in wind speed across different climatic and topographic zones. The network is stratified by elevation to ensure representation of coastal, mid-altitude, and interior wind regimes. 

\begin{figure}[hpbt]
	\centering
	\includegraphics[width=1.0\textwidth]{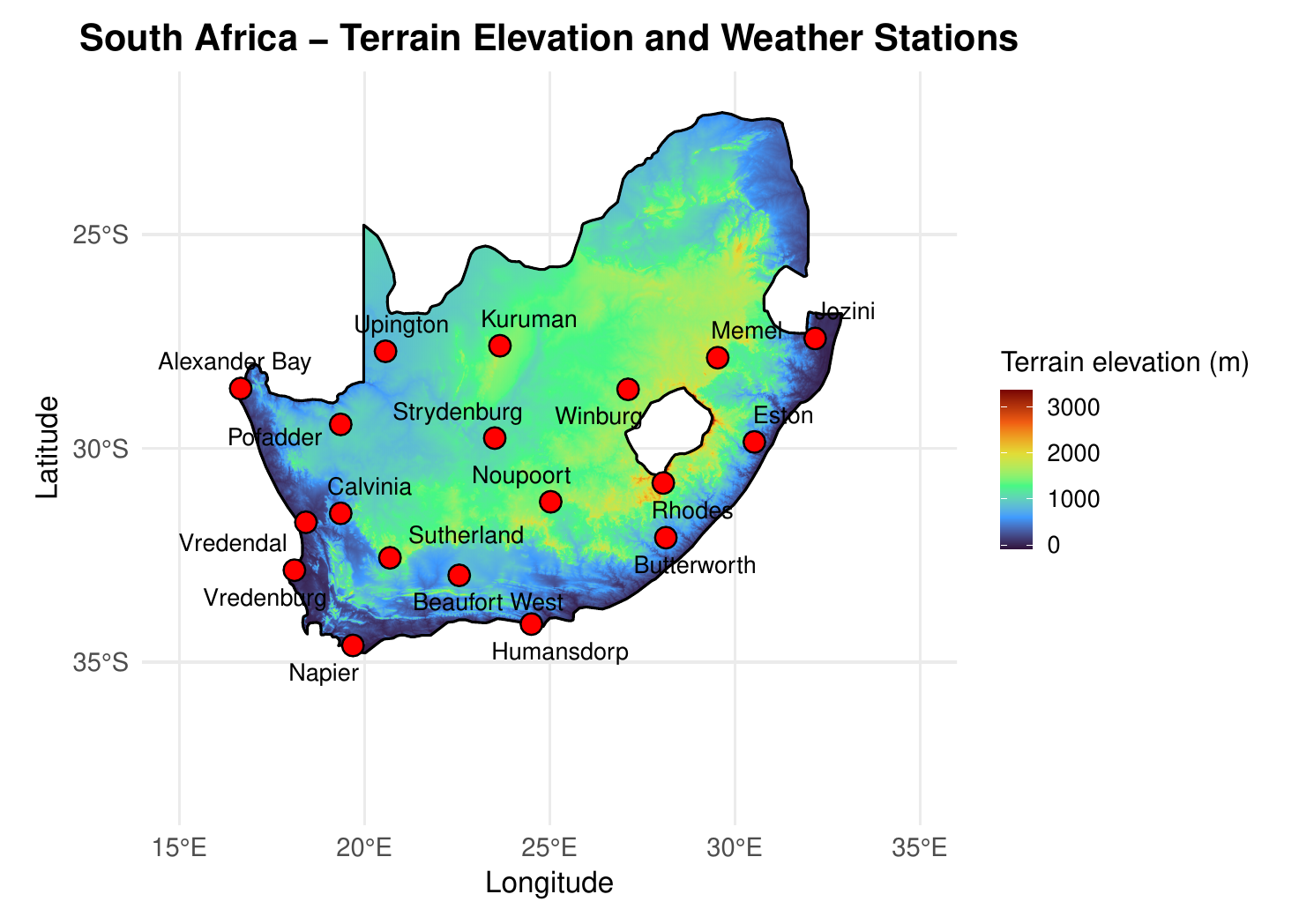}
	\caption{Spatial distribution of meteorological stations across South Africa.}
	\label{fig:station_map}
\end{figure}

Figure~\ref{fig:station_map} illustrates the stations' spatial distribution across South Africa. This spatial configuration enables the modelling of local dependence structures while accounting for regional-scale atmospheric drivers using the NNGP framework.

\subsection{Data Description}

Hourly meteorological data were collected over five years (2021-2025), resulting in approximately 43,800 observations per station. Although this high-frequency dataset provides a detailed spatio-temporal representation of near-surface atmospheric dynamics across heterogeneous terrain, the full hourly resolution was computationally infeasible for model estimation on the available hardware. Therefore, data was aggregated to daily averages to ensure computational tractability while preserving the dominant temporal dependence structure required for the proposed NNGP-AR(1) framework. 

Variable selection was carried out using Lasso to identify the most appropriate set of predictors of wind speed dynamics, resulting in a model that includes wind speed (m/s) as the response variable, the main variable of interest within the spatio-temporal setting. Several meteorological covariates (gust wind speed, air pressure, humidity and temperature) that reflect short-term atmospheric forcing, some spatial covariates such as elevation and cluster variables corresponding to mid-elevation and interior areas to address topography and regional differences, respectively, along with temporal persistence through the inclusion of first- and second-order lagged wind speed (wind\_lag1 and wind\_lag2), thus representing AR(1) type dependence. Seasonality is accounted for by introducing dummy variables for autumn, winter, and spring; finally, a time trend variable reflects long-term changes in wind speed behaviour.

\subsection{Data Preprocessing}

\subsubsection{Temporal Aggregation}

To reduce high-frequency noise and improve computational efficiency for the hierarchical Bayesian model, hourly observations were aggregated into daily means. Only days with at least 20 valid hourly observations were retained to ensure data reliability:

\[
\bar{w}_d = \frac{1}{n_h} \sum_{t=1}^{n_h} w_t, \quad n_h \geq 20
\]

This aggregation yields a final dataset of 19 stations over 1,818 days,
resulting in 34,542 observations for model estimation.

\subsubsection{Lagged Temporal Structure}

To explicitly capture temporal dependence, lagged wind speed variables were constructed:

\[
w_{t-1}, \quad w_{t-2}, \quad w_{t-7}
\]

These lags are designed to capture short-term persistence (daily memory) and weekly atmospheric cycles, thereby motivating the AR(1) temporal structure within the hierarchical model.

\section{Methodology} \label{sec:3}

\subsection*{Notation}
Let $Y(s,t)$ denote the observed wind speed at station location $s$ on day $t$, where $s = 1,\ldots,S$ indexes the 19 meteorological stations and $t = 1,\ldots,T$ indexes the time points in the study period. All models and forecasts are defined with respect to this fundamental quantity.

\paragraph{Flowchart of the process} 

\begin{figure}[H]
	\centering
	\adjustbox{max size={\textwidth}{\textheight}, center}{%
		\begin{tikzpicture}[
			node distance=0.8cm,          % tighter vertical spacing
			font=\small,                  % smaller font to save space
			startstop/.style = {rectangle, rounded corners, minimum width=4cm, minimum height=1.1cm, text centered, draw=black, fill=red!30, align=center},
			process/.style = {rectangle, minimum width=4cm, minimum height=1.1cm, text centered, draw=black, fill=blue!20, align=center},
			decision/.style = {diamond, minimum width=4cm, minimum height=1.3cm, text centered, draw=black, fill=green!30, align=center, aspect=1.2},
			arrow/.style = {thick,->,>=stealth}
			]
			
			% Nodes
			\node (start) [startstop] {Start:\\ Wind Speed Data $Y(s,t)$};
			\node (reg) [process, below=of start] {Step 1: Regression Baseline\\ Fit regression with covariates\\ Obtain residuals $e(s,t)$};
			\node (stationarity) [decision, below=of reg] {Residuals\\ stationary?};
			\node (ar1) [process, below=of stationarity] {Step 2: AR(1) Temporal Model\\ Fit per-station AR(1) to residuals\\ Generate forecasts $\widehat{Y}_{AR(1)}$};
			\node (spatial_res) [process, below=of ar1] {Step 3: Compute Spatial Residuals\\ $r(s,t)=Y(s,t)-\widehat{Y}_{AR(1)}(s,t)$};
			\node (nngp) [process, below=of spatial_res] {Step 4: NNGP Spatial Model\\ Fit NNGP with Matérn covariance\\ Estimate spatial effects $\widehat{w}(s)$};
			\node (hybrid) [process, below=of nngp] {Step 5: Hybrid Forecast\\ $\widehat{Y}_{Hybrid}=\widehat{Y}_{AR(1)}+\widehat{w}(s)$};
			\node (valid) [process, below=of hybrid] {Step 6: Validation\\ Rolling origin, RMSE, DM test,\\ Block conformal intervals};
			\node (final) [startstop, below=of valid] {Final Output:\\ Forecasts \& Intervals};
			
			% Arrows
			\draw [arrow] (start) -- (reg);
			\draw [arrow] (reg) -- (stationarity);
			\draw [arrow] (stationarity) -- node[anchor=east] {yes} (ar1);
			\draw [arrow] (stationarity.west) -- ++(-1.6,0) 
			node[anchor=south, pos=0.5, font=\normalsize] {no (revise)} 
			|- (reg.west);
			\draw [arrow] (ar1) -- (spatial_res);
			\draw [arrow] (spatial_res) -- (nngp);
			\draw [arrow] (nngp) -- (hybrid);
			\draw [arrow] (hybrid) -- (valid);
			\draw [arrow] (valid) -- (final);
			
		\end{tikzpicture}%
	}
	\caption{\small Flowchart of the structure of the~paper.}
	\label{flowchart}
\end{figure}
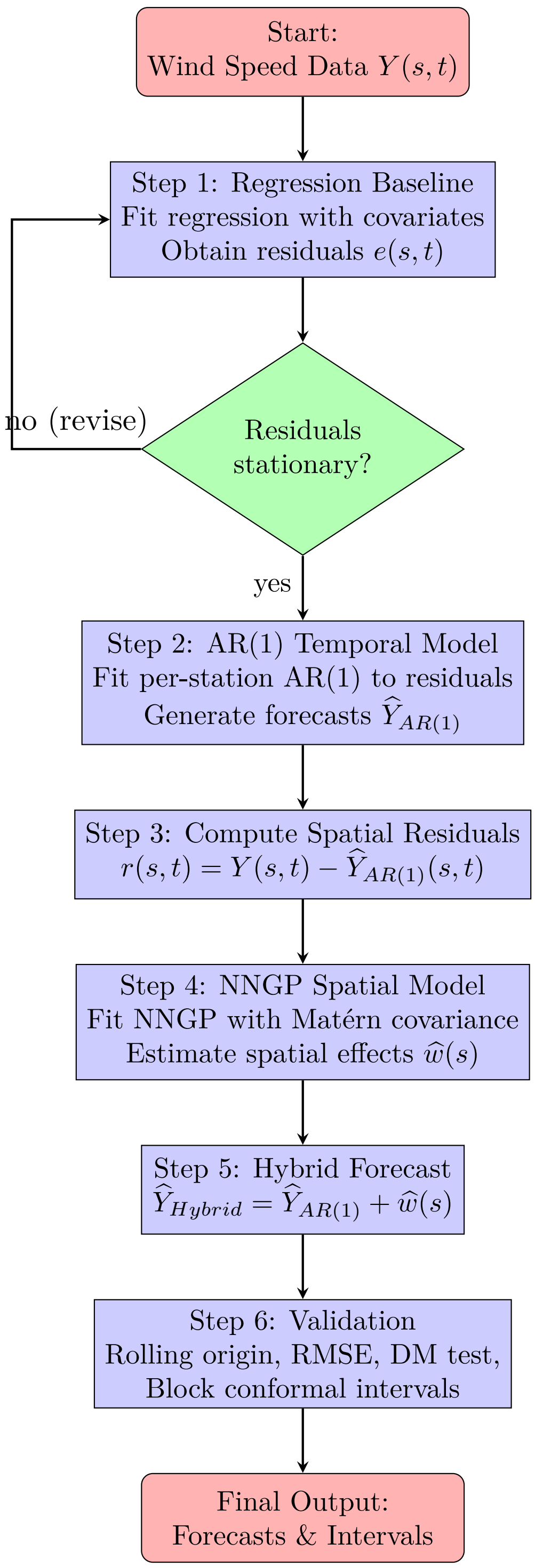

Figure \ref{flowchart} shows the flowchart of the structure of the paper. The flowchart outlines the steps of a hybrid approach for predicting wind speed. It begins with the input data, builds a regression model, and computes the residuals. There is an evaluation of stationarity at a decision point that takes a `o'  output to the loop to modify the model, while a ‘yes’ output proceeds to build a temporal model (AR(1)). After that, the spatial residuals are derived and passed to the NNGP model with the Matérn covariance function to build spatial models. The temporal and spatial components come together to build a hybrid prediction.

\subsection{Hierarchical Spatio-Temporal Modelling Framework}

The proposed framework decomposes wind speed forecasting into temporal and spatial components. The objective is to exploit temporal persistence within individual stations while capturing spatial dependence among stations through a scalable Gaussian process formulation. Let $Y(s,t)$ denote the observed wind speed at station location $s$ on day $t$. The proposed two-stage model is defined as:

\begin{equation}
	Y(s,t) = \widehat{Y}_{\text{AR(1)}}(s,t) + w(s) + \varepsilon(s,t),
	\label{eq:decomp}
\end{equation}

where:
\begin{itemize}
	\item $\widehat{Y}_{\text{AR(1)}}(s,t)$ represents the station-specific temporal forecast,
	\item $w(s)$ is a spatial Gaussian process capturing spatially-structured residual variation,
	\item $\varepsilon(s,t) \sim \mathcal{N}(0, \sigma^2_\varepsilon)$ is an independent error term.
\end{itemize}

In other words, the errors that remain to be forecast possess spatial structure that can be modelled independently of temporal dependence. The framework consists of four elements:
\begin{enumerate} 
	\item \textbf{Regression baseline} - to capture deterministic relationships between meteorological variables and wind speed, 
	\item \textbf{AR(1) temporal model} - to capture station-level temporal dependence, 
	\item \textbf{NNGP spatial residual model} - to capture remaining spatial structure, 
	\item \textbf{Hybrid forecast} - obtained by combining temporal and spatial components. 
\end{enumerate}

\subsection{Regression Baseline}

The initial model evaluates the contribution of meteorological and geographic covariates to wind speed variability. The regression model is specified as:

\begin{equation}
	Y(s,t) = \mathbf{x}(s,t)^\top \boldsymbol{\beta} + \varepsilon(s,t),
	\label{eq:reg}
\end{equation}
where $\mathbf{x}(s,t)$ contains covariates and $\boldsymbol{\beta}$ are regression coefficients. The covariate set includes wind gusts, pressure, humidity, temperature, lagged wind speeds ($Y_{t-1}, Y_{t-2}$), elevation, seasonal indicators, cluster membership, and a time trend. Fitted values provide a deterministic baseline:

\begin{equation}
	\widehat{Y}_{\text{reg}}(s,t) = \mathbf{x}(s,t)^\top \widehat{\boldsymbol{\beta}}.
	\label{eq:reg_fit}
\end{equation}

This model provides a deterministic benchmark before introducing temporal and spatial dependence structures.

\subsubsection{Elevation-Season Interaction}

To investigate whether the relationship between elevation and wind speed varies across seasons, the fixed-effects regression model was extended to include interaction terms between elevation and seasonal indicators. The model is specified as:

\begin{align}
	Y(s,t) = &\beta_0 + \beta_1 E(s) + \sum_{k=1}^{3} \gamma_k S_k(s,t) \nonumber \\
	&+ \sum_{k=1}^{3} \delta_k \{E(s) \times S_k(s,t)\} + \mathbf{x}(s,t)^\top \boldsymbol{\alpha} + \varepsilon(s,t),
	\label{eq:interaction}
\end{align}
where $E(s)$ is elevation and $S_k(s,t)$ are seasonal indicators (summer as reference). The interaction coefficients $\delta_k$ quantify whether the effect of elevation on wind speed changes across seasons. The null hypothesis $H_0: \delta_1 = \delta_2 = \delta_3 = 0$ is tested via F-test to assess the joint significance of elevation-season interactions.

Although exploratory analysis indicated pronounced seasonal variation in wind speed, a seasonal ARIMA structure was not introduced into the residual temporal model. Seasonal effects were explicitly represented in the regression component through seasonal indicator variables and their interactions with elevation.
Consequently, the AR(1) component was applied to the residual series after the deterministic seasonal structure had been removed. This specification separates systematic seasonal variation from short-term temporal persistence.
\subsection{Stationarity and Temporal Dependence Assessment}

\subsubsection{Stationarity Testing}

To validate the temporal modelling approach, we assess stationarity of the residual series using two complementary tests:

\begin{itemize}
	\item \textbf{Augmented Dickey-Fuller (ADF) Test}: Tests the null hypothesis that a unit root exists (non-stationary):
	\begin{equation}
		H_0: \rho = 1 \quad \text{vs.} \quad H_1: \rho < 1,
	\end{equation}
	where the test regression is:
	\begin{equation}
		\Delta e_t = \alpha + \rho e_{t-1} + \sum_{j=1}^{p} \psi_j \Delta e_{t-j} + \varepsilon_t.
	\end{equation}
	
	\item \textbf{KPSS Test}: Tests the null hypothesis that the series is stationary around a deterministic trend:
	\begin{equation}
		H_0: \sigma^2_\eta = 0 \quad \text{vs.} \quad H_1: \sigma^2_\eta > 0,
	\end{equation}
	with the decomposition:
	\begin{equation}
		e_t = \mu + \sum_{j=1}^{t} \eta_j + \varepsilon_t,
	\end{equation}
	where $\eta_j \sim (0, \sigma^2_\eta)$ and $\varepsilon_t \sim (0, \sigma^2_\varepsilon)$.
\end{itemize}

The KPSS test statistic is:
\begin{equation}
	\text{KPSS} = \frac{1}{T^2} \sum_{t=1}^{T} \frac{S_t^2}{\hat{\sigma}^2_\varepsilon},
	\label{eq:kpss}
\end{equation}
where $S_t = \sum_{j=1}^{t} \hat{\varepsilon}_j$ are partial sums of residuals.

\subsubsection{Temporal Dependence Structure}

The temporal autocorrelation structure is examined via the autocorrelation function (ACF):
\begin{equation}
	\rho(h) = \frac{\text{Cov}(e_t, e_{t-h})}{\sqrt{\text{Var}(e_t)\text{Var}(e_{t-h})}},
	\label{eq:acf}
\end{equation}
and the partial autocorrelation function (PACF), which measures the correlation between $e_t$ and $e_{t-h}$ after removing the effects of intermediate lags.

The optimal autoregressive order is determined using Akaike's Information Criterion corrected for small samples (AICc):
\begin{equation}
	\text{AICc} = -2\log\mathcal{L}(\hat{\boldsymbol{\theta}}) + 2p + \frac{2p(p+1)}{T - p - 1},
	\label{eq:aicc}
\end{equation}
where $\mathcal{L}(\hat{\boldsymbol{\theta}})$ is the maximized likelihood, $p$ is the number of parameters, and $T$ is the sample size.

\subsection{AR(1) Temporal Component}

For each station independently, the residual series from the regression model was modelled using an autoregressive process. The temporal component is given by:

\begin{equation}
	e(s,t) = \phi(s) \, e(s,t-1) + \eta(s,t),
	\label{eq:ar1}
\end{equation}
with $\eta(s,t) \sim \mathcal{N}(0, \sigma^2_\eta(s))$. The global autoregressive parameter is estimated as:
\begin{equation}
	\hat{\phi} = \frac{1}{S}\sum_{s=1}^{S} \hat{\phi}(s).
	\label{eq:global_phi}
\end{equation}

The $h$-step-ahead forecast:

\begin{equation}
	\widehat{Y}_{\text{AR(1)}}(s,t+h) = \widehat{Y}_{\text{reg}}(s,t+h) + \phi(s)^h e(s,t).
	\label{eq:arima_forecast}
\end{equation}

The AR(1) forecasts therefore account for temporal persistence and short-term station-specific dynamics.

\subsection{Spatial Residual Process}
After AR(1) forecasting, spatial residuals are:

\begin{equation}
	r(s,t) = Y(s,t) - \widehat{Y}_{\text{ARIMA}}(s,t).
	\label{eq:spatial_resid}
\end{equation}

These are modelled via a Gaussian process:

\begin{equation}
	w(s) \sim \mathcal{GP}\left(\mu, C(s,s'; \boldsymbol{\theta})\right),
	\label{eq:gp}
\end{equation}

with Matérn covariance:

\begin{equation}
	C(s,s') = \sigma^2 \rho(h; \phi, \nu) + \tau^2 \mathbbm{1}(s = s'),
	\label{eq:matern}
\end{equation}

where $h = \|s - s'\|$ is the distance between stations,, $\sigma^2$ is spatial variance, $\tau^2$ is nugget variance, $\phi$ controls spatial decay, and $\nu$ controls smoothness. The Matérn correlation function is:

\begin{equation}
	\rho(h; \phi, \nu) = \frac{1}{2^{\nu-1}\Gamma(\nu)}(\phi h)^\nu K_\nu(\phi h),
	\label{eq:matern_corr}
\end{equation}

where $K_\nu(\cdot)$ is the modified Bessel function of the second kind. The spatial range is:
\begin{equation}
	R = \frac{3}{\phi} \times 111 \text{ km},
	\label{eq:range}
\end{equation}
where 111 km converts degrees to kilometers.

\subsection{Nearest Neighbour Gaussian Process Approximation}

Because Gaussian process inference scales cubically with the number of spatial locations, a Nearest Neighbour Gaussian Process (NNGP) approximation was used. For station location $s_i$, the conditional distribution was approximated as:
\begin{equation}
	w(s_i) \mid w(s_{-i}) \approx w(s_i) \mid w(\mathcal{N}(s_i)),
	\label{eq:nngp}
\end{equation}

where $\mathcal{N}(s_i)$ denotes the set of $m$ nearest neighbours of location $s_i$. The neighbour set is defined by Euclidean distance:
\begin{equation}
	\mathcal{N}(s_i) = \left\{ s_j : \|s_i - s_j\| \text{ is among the } m \text{ smallest distances} \right\}.
	\label{eq:neighbours}
\end{equation}

The NNGP likelihood factorization is:
\begin{equation}
	p(\mathbf{w}) \approx \prod_{i=1}^{S} p\left(w(s_i) \mid \{w(s_j): s_j \in \mathcal{N}(s_i)\}\right).
	\label{eq:nngp_lik}
\end{equation} Eight nearest neighbours were used in the spatial approximation. This produces a sparse representation of the spatial covariance structure while retaining the dominant spatial dependence among stations.

\subsection{Hybrid Forecast}

This forecasting model will then be the combination of the temporal forecast and the spatial residual estimate:

\begin{equation}
	\widehat{Y}_{\text{Hybrid}}(s,t+h) = \widehat{Y}_{\text{AR(1)}}(s,t+h) + \widehat{w}(s),
	\label{eq:hybrid}
\end{equation}

where $\widehat{w}(s)$ represents the posterior mean for the spatial residual effect at each station level using the NNGP approach.

The above formula decomposes the data into temporal and spatial components, assuming that temporal dynamics and spatial bias are independent sources of variation. The former accounts for changes in winds, while the latter corrects for systematic station-level biases.

\subsubsection{Covariate-Enhanced NNGP Spatial Model}

To investigate whether known geographic characteristics could further explain spatial variation, an extended NNGP model incorporating elevation and terrain clusters was evaluated. The spatial residual process was modified to include spatial covariates:

\begin{equation}
	r(s,t) = \mathbf{X}(s)\boldsymbol{\beta} + w(s) + \xi(s,t),
	\label{eq:cov_spatial}
\end{equation}
where $\mathbf{X}(s) = (E(s), C_1(s), C_2(s))$ includes elevation and cluster indicators (coastal as reference). The enhanced hybrid forecast:

\begin{equation}
	\widehat{Y}_{\text{Hybrid}}(s,t+h) = \widehat{Y}_{\text{AR(1)}}(s,t+h) + \mathbf{X}(s)\widehat{\boldsymbol{\beta}} + \widehat{w}(s).
	\label{eq:enhanced_hybrid}
\end{equation}

This model was evaluated to determine whether explicitly modelling geographic characteristics improved upon the latent spatial correction provided by the NNGP.

\subsection{Horizon-Dependent Spatial Correction}

An additional sensitivity analysis was conducted to examine whether the contribution of spatial information decreases with lead time. The alternative formulation with decay parameter was:

\begin{equation}
	\widehat{Y}_{\text{Hybrid}}(s,t+h) = \widehat{Y}_{\text{AR(1)}}(s,t+h) + \lambda_h \widehat{w}(s),
	\label{eq:decay}
\end{equation}

where:

\begin{equation}
	\lambda_h = \exp(-\rho h),
	\label{eq:lambda}
\end{equation}
where $\lambda_h = \exp(-\rho h)$ and $\rho$ controls temporal decay rate. The optimal decay rate is found by minimizing RMSE over a grid of $\rho$ values.

\subsection{Bayesian Parameter Estimation and MCMC Implementation}

The NNGP model was estimated using Bayesian inference implemented through Markov chain Monte Carlo (MCMC) sampling. The prior distributions were specified as:

\begin{align}
	\sigma^2 &\sim \mathcal{IG}(2, 1), \label{eq:prior_sigma} \\
	\tau^2 &\sim \mathcal{IG}(2, 0.5), \label{eq:prior_tau} \\
	\phi &\sim \mathcal{U}(0.5, 10), \label{eq:prior_phi} \\
	\nu &\sim \mathcal{U}(0.5, 2). \label{eq:prior_nu}
\end{align}

The inverse gamma priors provide weakly informative distributions for variance components, while the uniform priors constrain the spatial decay and smoothness parameters to physically plausible ranges. The decay parameter range was chosen to allow both short-range and regional-scale spatial dependence across the South African station network.
Posterior samples are obtained via Markov Chain Monte Carlo (MCMC) with 30,000 iterations and a burn-in period of 15,000 samples. Convergence is assessed using the Gelman-Rubin diagnostic:
\begin{equation}
	\hat{R} = \sqrt{\frac{\widehat{\text{Var}}(\theta)}{W}},
	\label{eq:gelman}
\end{equation}
where $W$ is the within-chain variance and $\widehat{\text{Var}}(\theta)$ is the estimated posterior variance. Values of $\hat{R} < 1.1$ indicate convergence.

\subsection{Forecast Validation}

Forecast performance was assessed using a rolling-origin evaluation procedure designed to mimic operational forecasting. Five forecast origins were selected throughout the 1,818-day study period, with the first forecast origin occurring after 100 days of available observations. At each forecast origin, the observations available up to that point were divided into a proper training period and a calibration period, with approximately 20\% of the available observations allocated to calibration and the remaining observations used for model estimation. Ten consecutive one-day-ahead forecasts ($h = 1,\ldots,10$) were then generated, providing a 10-day out-of-sample evaluation period at each forecast origin. 

The validation design is summarised in Table~\ref{tab:rolling_origin}.

\begin{table}[h!]
	\centering
	\caption{Rolling-origin validation design.}
	\label{tab:rolling_origin}
	\begin{tabular}{|c|c|c|c|c|}
		\hline
		\textbf{Origin} & \textbf{Training Days} & \textbf{Calibration Days} & \textbf{Forecast Origin} & \textbf{Test Days} \\
		\hline
		1 & 1--80 & 81--100 & 100 & 101--110 \\
		2 & 1--422 & 423--527 & 527 & 528--537 \\
		3 & 1--764 & 765--954 & 954 & 955--964 \\
		4 & 1--1105 & 1106--1381 & 1381 & 1382--1391 \\
		5 & 1--1447 & 1448--1808 & 1808 & 1809--1818 \\
		\hline
	\end{tabular}
\end{table}
The expanding training windows allow progressively more historical observations to be incorporated as the forecast origin advances, thereby providing a validation framework that approximates operational forecasting conditions.
For each horizon $h$, forecasts were compared between the AR(1) baseline and the proposed AR(1)–NNGP model. The primary evaluation metric was root mean square error:

\begin{equation}
	\text{RMSE} = \sqrt{\frac{1}{n} \sum_{i=1}^{n} \bigl(Y_i - \widehat{Y}_i\bigr)^2}.
	\label{eq:rmse}
\end{equation}

\subsection{Block-Wise Conformal Prediction Intervals}

\subsubsection{Calibration and Block Construction}

Prediction uncertainty for the final AR(1)-NNGP hybrid forecasts was quantified using block-wise conformal prediction. Let $S=19$ denote the number of meteorological stations and let $T_{\mathrm{cal}}$ denote the number of calibration days. For each station $s=1,\ldots,S$ and calibration day $t=1,\ldots,T_{\mathrm{cal}}$, the absolute calibration error was calculated as

\begin{equation}
	e(s,t)
	=
	\left|
	Y(s,t)
	-
	\widehat{Y}_{\mathrm{Hybrid}}(s,t)
	\right|,
	\label{eq:calib_error}
\end{equation}

where $Y(s,t)$ denotes the observed wind speed at station $s$ on day $t$, and $\widehat{Y}_{\mathrm{Hybrid}}(s,t)$ denotes the corresponding AR(1)-NNGP hybrid forecast. These absolute errors serve as the nonconformity scores used to construct the conformal prediction intervals.

To account for temporal dependence in the calibration errors, the calibration period was partitioned into non-overlapping blocks of $B$ consecutive days. For a given block size $B$, the $k$th block is defined as

\begin{equation}
	\mathcal{T}_k
	=
	\left\{
	t_k,\ldots,t_k+B-1
	\right\},
	\qquad
	k=1,\ldots,K,
	\label{eq:blocks}
\end{equation}

where, $K = \left\lfloor \frac{T_{\mathrm{cal}}}{B} \right\rfloor$ is the number of complete calibration blocks. If any calibration days remain that do not make up an entire block, they are not used to calculate the block score.

For each block, a conservative block score was calculated as the maximum calibration error of all sites in the block:

\begin{equation}
	s_k
	=
	\max_{s=1,\ldots,S}
	\max_{t\in\mathcal{T}_k}
	e(s,t),
	\qquad
	k=1,\ldots,K.
	\label{eq:block_score}
\end{equation}

Therefore, each block has a single score: the maximum calibration error within that block of time and space. Such an approach results in a uniform error threshold that accounts for both temporal dependencies within blocks and spatial dependencies between stations.
\subsubsection{Block Size Selection}

Several block sizes were investigated to assess the effect of temporal aggregation on the empirical coverage and sharpness of the conformal prediction intervals. The candidate block sizes were $B=\{3,5,7,8,9\}$.
For each candidate block size $B$, the calibration period was partitioned into non-overlapping blocks and block-level nonconformity scores were calculated. Prediction intervals were then constructed using the corresponding empirical conformal quantile and evaluated on the out-of-sample observations.

The empirical coverage for a given block size was calculated as

\begin{equation}
	\widehat{\mathrm{Coverage}}(B)
	=
	\frac{1}{S H}
	\sum_{s=1}^{S}
	\sum_{h=1}^{H}
	\mathbb{I}
	\left\{
	Y(s,t+h)
	\in
	\mathrm{PI}_{1-\alpha}^{(B)}(s,t+h)
	\right\},
	\label{eq:emp_coverage}
\end{equation}

where $S=19$ denotes the number of stations and $H$ denotes the forecast horizon.

A nominal miscoverage level of $\alpha=0.10$ was used, corresponding to a target coverage of $90\%$. 

\subsubsection{Prediction Interval Construction}

The block-level conformal quantile was obtained as the empirical $(1-\alpha)$ quantile of the $K$ block scores:

\begin{equation}
	\widehat{q}_{1-\alpha}
	=
	\inf
	\left\{
	q:
	\frac{1}{K}
	\sum_{k=1}^{K}
	\mathbb{I}(s_k\leq q)
	\geq
	1-\alpha
	\right\}.
	\label{eq:block_quantile}
\end{equation}

For a new hybrid forecast at station $s$ and forecast horizon $h$, the corresponding two-sided conformal prediction interval is given by

\begin{equation}
	\mathrm{PI}_{1-\alpha}(s,t+h)
	=
	\left[
	\widehat{Y}_{\mathrm{Hybrid}}(s,t+h)
	-
	\widehat{q}_{1-\alpha},
	\;
	\widehat{Y}_{\mathrm{Hybrid}}(s,t+h)
	+
	\widehat{q}_{1-\alpha}
	\right].
	\label{eq:pred_interval_block}
\end{equation}

Because wind speed is physically non-negative, the lower bound was truncated at zero. The final prediction interval was therefore defined as

\begin{equation}
	\mathrm{PI}_{1-\alpha}(s,t+h)
	=
	\left[
	\max
	\left\{
	0,
	\widehat{Y}_{\mathrm{Hybrid}}(s,t+h)
	-
	\widehat{q}_{1-\alpha}
	\right\},
	\;
	\widehat{Y}_{\mathrm{Hybrid}}(s,t+h)
	+
	\widehat{q}_{1-\alpha}
	\right].
	\label{eq:pred_interval_bound}
\end{equation}

The resulting intervals provide distribution-free uncertainty quantification around the AR(1)-NNGP hybrid forecasts while accounting for temporal dependence through the block construction.

\subsection{Diebold-Mariano Test}

The predictive accuracy of the AR(1) benchmark and the proposed AR(1)-NNGP hybrid model was compared using the Diebold--Mariano (DM) test \citep{diebold1995comparing}. The test was applied to determine whether differences in predictive accuracy between the two models were statistically significant at each forecast horizon.

For station $s$ and forecast horizon $h$, the forecast errors for the AR(1) and hybrid models were defined as

\begin{equation}
	e_{\mathrm{AR(1)}}(s,h)
	=
	Y(s,t+h)
	-
	\widehat{Y}_{\mathrm{AR(1)}}(s,t+h),
	\label{eq:dm_arima_error}
\end{equation}

and

\begin{equation}
	e_{\mathrm{Hybrid}}(s,h)
	=
	Y(s,t+h)
	-
	\widehat{Y}_{\mathrm{Hybrid}}(s,t+h),
	\label{eq:dm_hybrid_error}
\end{equation}

where $Y(s,t+h)$ denotes the observed wind speed at station $s$ and forecast horizon $h$, while $\widehat{Y}_{\mathrm{AR(1)}}(s,t+h)$ and $\widehat{Y}_{\mathrm{Hybrid}}(s,t+h)$ denote the corresponding forecasts from the AR(1) and AR(1)-NNGP hybrid models.

Squared-error loss was used in the DM test. The loss differential was therefore defined as

\begin{equation}
	d(s,h)
	=
	e_{\mathrm{AR(1)}}^2(s,h)
	-
	e_{\mathrm{Hybrid}}^2(s,h).
	\label{eq:loss_diff}
\end{equation}

Under this definition, a positive loss differential indicates that the hybrid model has a smaller squared forecast error than the AR(1) model.

The null hypothesis of equal predictive accuracy is

\begin{equation}
	H_0:
	\mathbb{E}[d(s,h)] = 0,
	\label{eq:dm_h0}
\end{equation}

while the two-sided alternative hypothesis is

\begin{equation}
	H_1:
	\mathbb{E}[d(s,h)] \neq 0.
	\label{eq:dm_h1}
\end{equation}

The DM statistic is given by

\begin{equation}
	\mathrm{DM}
	=
	\frac{\bar{d}}
	{\sqrt{\widehat{\mathrm{Var}}(\bar{d})}},
	\label{eq:dm_stat}
\end{equation}

where $\bar{d}$ is the mean loss differential and $\widehat{\mathrm{Var}}(\bar{d})$ is its estimated variance.

The DM test was conducted separately for each forecast horizon, $h=1,\ldots,10$. At each horizon, the forecast errors from the 19 meteorological stations were compared between the AR(1) and AR(1)-NNGP hybrid models. In addition to the DM statistic and corresponding $p$-value, the root mean squared error (RMSE) was calculated for each model:

\begin{equation}
	\mathrm{RMSE}_{k,h}
	=
	\sqrt{
		\frac{1}{S}
		\sum_{s=1}^{S}
		e_k^2(s,h)
	},
	\qquad
	k\in\{\mathrm{ARIMA},\mathrm{Hybrid}\},
	\label{eq:dm_rmse}
\end{equation}

where $S=19$ is the number of meteorological stations. The forecast error magnitude, measured by horizon-specific RMSE values, indicates the magnitude of forecast error. At the same time, the DM test determines whether there is a significant difference in predictive accuracy between the two models.

Finally, we applied a pooled DM test, combining the forecast errors of all the considered stations and forecast horizons. This provided one test of the overall predictive performance of the AR(1) and AR(1)-NNGP model hybrids over the entire forecast evaluation period.

\section{Results} \label{sec:4}
\subsection{Exploratory data analysis}
\begin{figure}[!htbp]
	\centering
	\includegraphics[width=0.8\textwidth]{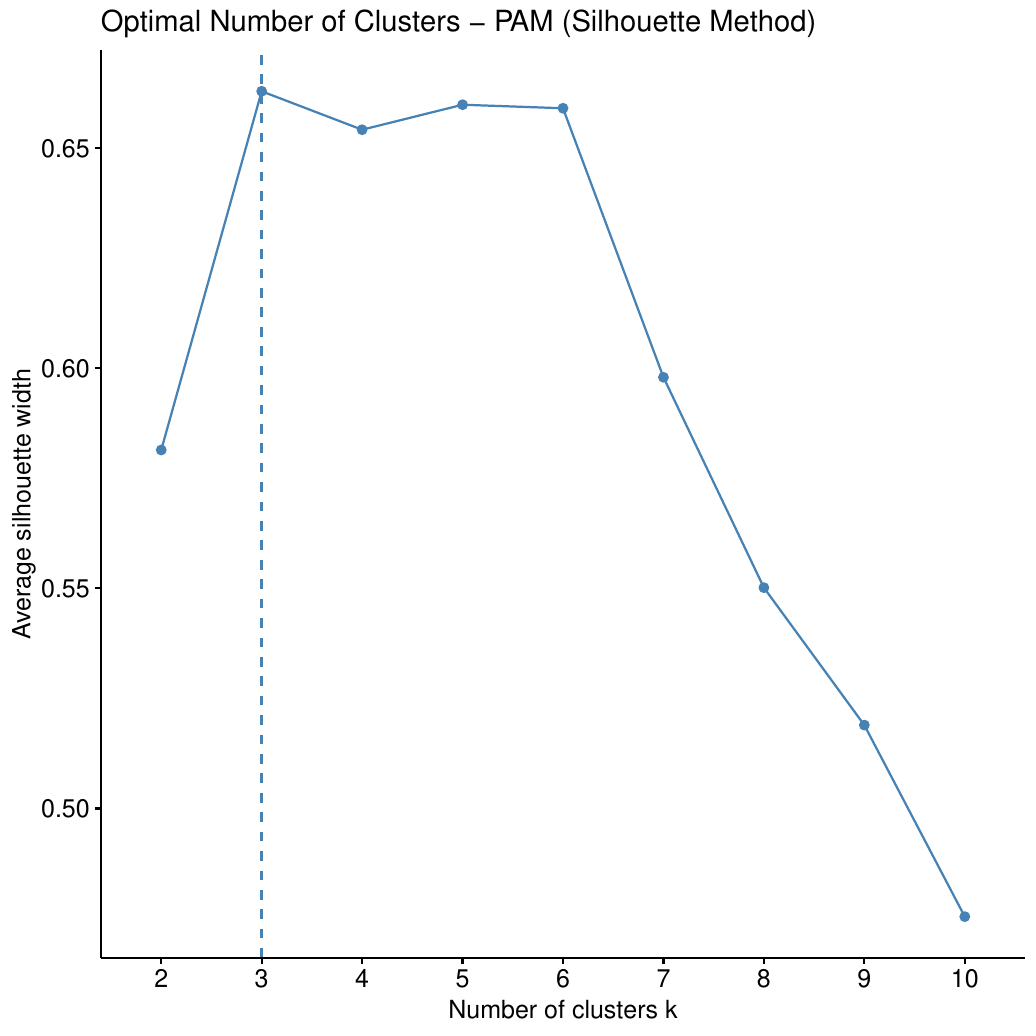}
	\caption{Silhoutte plot showing the optimal number of clusters.}
	\label{fig:elbow}
\end{figure}

\begin{figure}[!htbp]
	\centering
	\includegraphics[width=0.8\textwidth]{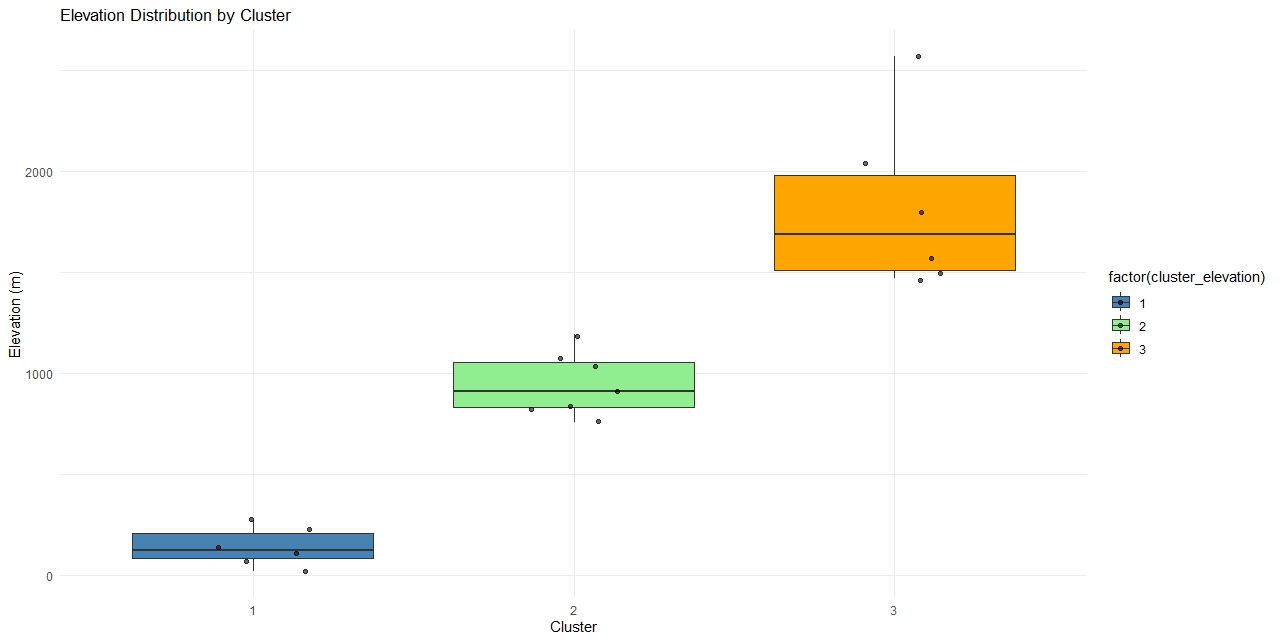}
	\caption{Boxplot showing station clusters by elevation across South Africa.}
	\label{fig:boxplot}
\end{figure}

\paragraph{Clustering}
Clusters were selected based on Partitioning Around Medoids, often referred to as K-Medoids. We used the Silhouette approach for selecting the number of clusters. From the clustering results, it is evident that there are three groups of elevations with 6, 7, and 6 stations each. Figure~\ref{fig:elbow} supports the selection of three optimal clusters. Box plots of the elevations of the stations are shown in Figure~\ref{fig:boxplot}.

\FloatBarrier

Table \ref{tab:clusters} provides the complete list of monitoring stations.

\begin{table}[htbp]
	\centering
	\scriptsize
	\setlength{\tabcolsep}{2.5pt}
	\caption{Station clusters and summary of cluster characteristics.}
	\label{tab:clusters}
	
	\begin{tabular}{@{}%
			p{0.07\textwidth}
			p{0.19\textwidth}
			c
			c
			c
			c
			@{}}
		\toprule
		StationNo & Station &
		Longitude ($^\circ$) &
		Latitude ($^\circ$) &
		Elevation (m) &
		Cluster \\
		\midrule
		
		WM1  & Alexander Bay        & 16.7 & -28.6 & 152  & 1 \\
		WM8  & Humansdorp (medoid)  & 24.5 & -34.1 & 110  & 1 \\
		WM13 & Jozini               & 32.2 & -27.4 & 80   & 1 \\
		WM5  & Napier               & 19.7 & -34.6 & 288  & 1 \\
		WM4  & Vredenburg           & 18.1 & -32.8 & 22   & 1 \\
		WM3  & Vredendal            & 18.4 & -31.7 & 242  & 1 \\
		
		\addlinespace
		
		WM7  & Beaufort West        & 22.6 & -33.0 & 1047 & 2 \\
		WM10 & Butterworth (medoid) & 28.1 & -32.1 & 925  & 2 \\
		WM2  & Calvinia             & 19.4 & -31.5 & 824  & 2 \\
		WM12 & Eston                & 30.5 & -29.8 & 770  & 2 \\
		WM16 & Pofadder             & 19.4 & -29.4 & 1073 & 2 \\
		WM17 & Strydenburg          & 23.5 & -29.8 & 1196 & 2 \\
		WM19 & Upington             & 20.6 & -27.7 & 848  & 2 \\
		
		\addlinespace
		
		WM18 & Kuruman              & 23.7 & -27.6 & 1478 & 3 \\
		WM14 & Memel                & 29.5 & -27.9 & 2045 & 3 \\
		WM9  & Noupoort (medoid)    & 25.0 & -31.2 & 1806 & 3 \\
		WM11 & Rhodes               & 28.1 & -30.8 & 2575 & 3 \\
		WM6  & Sutherland           & 20.7 & -32.6 & 1581 & 3 \\
		WM15 & Winburg              & 27.1 & -28.6 & 1505 & 3 \\
		
		\midrule
		\multicolumn{6}{l}{\textbf{Summary of clusters}} \\
		\cmidrule(lr){1-6}
		
		\multicolumn{2}{l}{Cluster} &
		\multicolumn{1}{c}{Lon. range ($^\circ$)} &
		\multicolumn{1}{c}{Lat. range ($^\circ$)} &
		\multicolumn{1}{c}{Elev. range (m)} &
		\multicolumn{1}{c}{No. stations} \\
		
		\cmidrule(lr){1-6}
		
		\multicolumn{2}{l}{Cluster 1} &
		16.66--32.17 &
		$-34.61$ to $-27.43$ &
		22--288 &
		6 \\
		
		\multicolumn{2}{l}{Cluster 2} &
		19.36--30.53 &
		$-32.97$ to $-27.73$ &
		770--1196 &
		7 \\
		
		\multicolumn{2}{l}{Cluster 3} &
		20.69--29.54 &
		$-32.56$ to $-27.60$ &
		1478--2575 &
		6 \\
		
		\bottomrule
		\multicolumn{6}{@{}l}{\footnotesize
			Coordinates are in decimal degrees; elevation is in metres.}
	\end{tabular}
\end{table}

\begin{figure}[hpbt]
	\centering
	\includegraphics[width=0.8\textwidth]{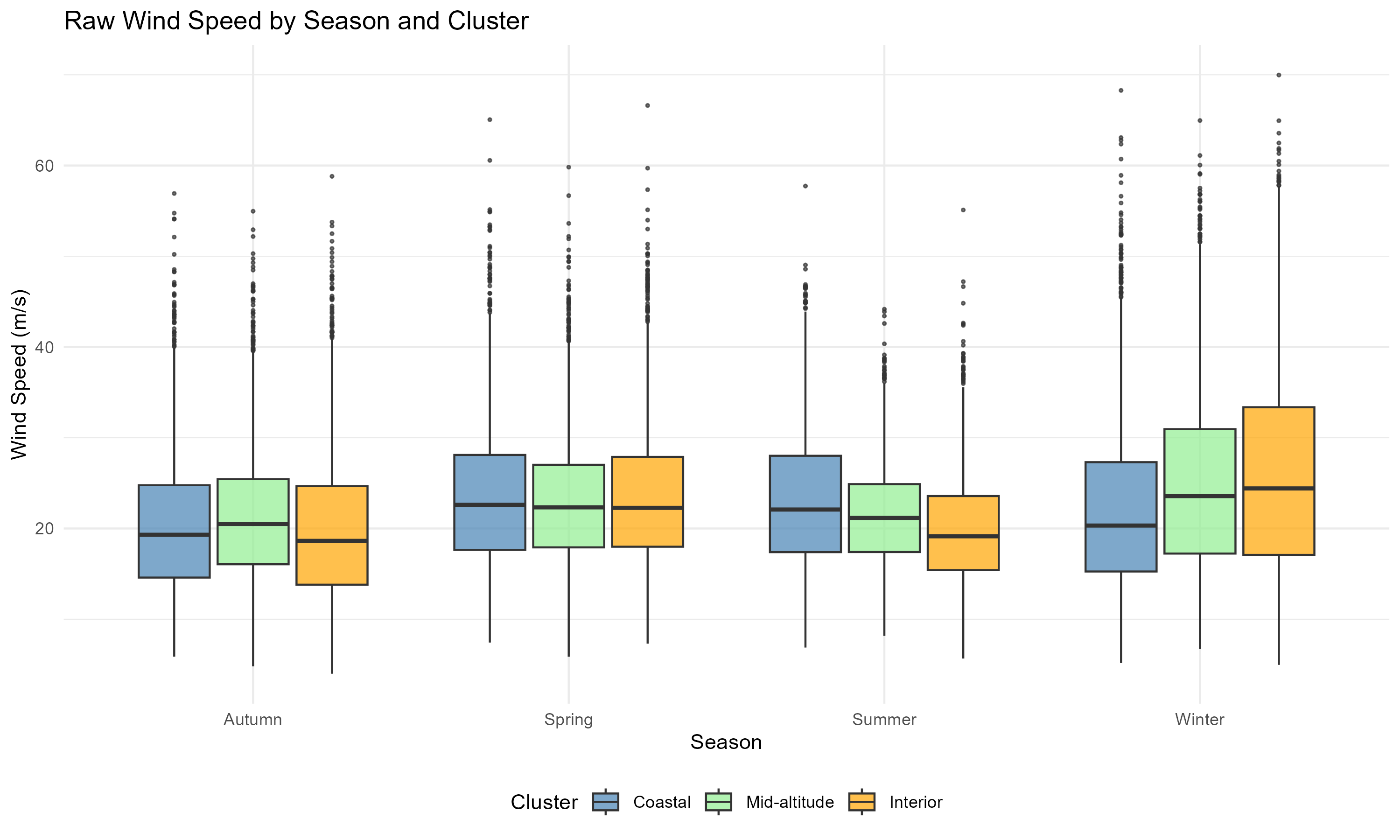}
	\caption{Boxplot showing station clusters by season across South Africa.}
	\label{fig:raw wind}
\end{figure}

\paragraph{Raw wind speed by season}
Figure~\ref{fig:raw wind} illustrates the distribution of raw wind speed (m/s) across four seasons (Autumn, Spring, Summer, and Winter), categorized by three geographic clusters: Coastal (blue), Mid-altitude (green), and Interior (orange).

\subsection{Fixed-Effects Regression Results}

The estimated coefficients of the fixed-effects regression model are reported in Table~\ref{tab:regression_results}. The model includes meteorological variables, geographical characteristics, seasonal indicators, lagged wind speed, a temporal trend and interaction terms between elevation and season.
\begin{table}[ht]
	\centering
	\caption{Fixed-effects regression model with elevation--season interactions.}
	\label{tab:regression_results}
	\begin{tabular}{lrrrr}
		\hline
		Variable & Estimate & Std. Error & $t$-value & $p$-value \\
		\hline
		Intercept & $-90.220$ & 3.672 & $-24.574$ & $<0.001$ \\
		Wind gusts & $0.7934$ & 0.0016 & 497.182 & $<0.001$ \\
		Pressure & $0.0903$ & 0.0036 & 25.360 & $<0.001$ \\
		Humidity & $-0.0513$ & 0.0009 & $-57.490$ & $<0.001$ \\
		Elevation & $-0.004844$ & 0.00007 & $-73.009$ & $<0.001$ \\
		Autumn & $2.036$ & 0.065 & 31.343 & $<0.001$ \\
		Winter & $3.650$ & 0.072 & 50.880 & $<0.001$ \\
		Spring & $0.827$ & 0.066 & 12.526 & $<0.001$ \\
		Temperature & $-0.0146$ & 0.0045 & $-3.247$ & 0.001 \\
		Wind lag$_1$ & $0.0168$ & 0.0020 & 8.310 & $<0.001$ \\
		Wind lag$_2$ & $-0.0047$ & 0.0018 & $-2.555$ & 0.011 \\
		Mid cluster & $4.887$ & 0.058 & 83.951 & $<0.001$ \\
		Interior cluster & $6.159$ & 0.101 & 61.104 & $<0.001$ \\
		Time trend & $1.511$ & 0.047 & 32.264 & $<0.001$ \\
		Elevation $\times$ Autumn & $-0.000268$ & 0.000053 & $-5.065$ & $<0.001$ \\
		Elevation $\times$ Winter & $-0.001034$ & 0.000054 & $-18.984$ & $<0.001$ \\
		Elevation $\times$ Spring & $-0.000587$ & 0.000053 & $-11.067$ & $<0.001$ \\
		\hline
	\end{tabular}
\end{table}

The regression model explained 91.3\% of the variance in wind speed, with the coefficient of determination $R^{2}=0.9129$ and the adjusted $R^{2}=0.9129$. The regression analysis was statistically significant for the full model ($p<0.001$). Thus, the set of chosen explanatory variables provides a good description of the deterministic part of wind speed.
Among all predictors, wind gust had the highest positive correlation with wind speed ($\hat{\beta}=0.7934$, $p<0.001$). This means that if there is a change of 1 unit in wind gust, the average change in wind speed is 0.79 m/s, ceteris paribus.
Besides, atmospheric pressure had a positive and statistically significant relationship with wind speed ($\hat{\beta}=0.0903$, $p<0.001$), while relative humidity had a negative and statistically significant relationship ($\hat{\beta}=-0.0513$, $p<0.001$).

Elevation had a significant negative main effect ($\hat{\beta}=-0.004844$, $p<0.001$), indicating that, during the reference season (summer), wind speed decreases by approximately $0.004844$~m/s for every one-metre increase in elevation. However, this effect varied significantly across seasons.

All the seasonal dummies were statistically significant. Compared to summer, the average wind speed increased by approximately $2.04$~m/s in autumn, by $3.65$~m/s in winter, and by $0.83$~m/s in spring.

The temperature variable showed a statistically significant but very small negative effect on the dependent variable ($\hat{\beta}=-0.0146$, $p=0.001$). The lagged wind speed variables had significant effects in the model, indicating some persistence in wind speed.

Both geographical cluster indicators were highly significant. Compared with the coastal reference cluster, stations located in the mid-region experienced average wind speeds approximately $4.89$~m/s higher, while stations in the interior experienced average wind speeds approximately $6.16$~m/s higher after accounting for the remaining covariates.

\subsubsection{Elevation-Season Interaction Effects}

To determine whether the influence of elevation changes across seasons, interaction terms between elevation and seasonal indicators were incorporated into the regression model.

The estimated interaction coefficients were
\begin{align}
	\widehat{\beta}_{E \times \text{Autumn}} &= -0.000268, \\
	\widehat{\beta}_{E \times \text{Winter}} &= -0.001034, \\
	\widehat{\beta}_{E \times \text{Spring}} &= -0.000587.
\end{align}

All interaction coefficients were highly statistically significant ($p<0.001$), indicating that the relationship between elevation and wind speed is season dependent.
Consequently, the marginal effect of elevation differs by season. The resulting seasonal effects are shown in Table~\ref{tab:marginal}.

\begin{table}[htbp]
	\centering
	\caption{Marginal effect of elevation on wind speed by season.}
	\label{tab:marginal}
	\begin{tabular}{lc}
		\hline
		Season & Marginal effect (m/s per metre)\\
		\hline
		Summer (baseline) & $-0.004844$ \\
		Autumn & $-0.005112$ \\
		Winter & $-0.005878$ \\
		Spring & $-0.005431$ \\
		\hline
	\end{tabular}
\end{table}

The strongest negative effect of elevation occurred during winter, where each one-metre increase in elevation reduced average wind speed by approximately $0.00588$~m/s. Autumn and spring also exhibited stronger negative elevation effects than summer, although the winter effect was substantially larger.
To assess whether the interaction terms jointly improved the regression model, an overall $F$-test was performed. The test yielded

\[
F = 134.78,\qquad p < 0.001,
\]
providing overwhelming evidence that the elevation-season interaction terms are jointly significant. Consequently, the interaction model provides a substantially better description of the deterministic component of wind speed than a model assuming a constant elevation effect throughout the year.
These findings justify retaining the interaction terms in the deterministic regression model before proceeding to temporal modelling using ARIMA and subsequent spatial modelling using the NNGP.

\subsubsection{Seasonal and spatial effects}
After performing regression analysis, descriptive and model-based summaries were used to understand the effects of seasons, elevation, and space on wind speed. Observed mean wind speed according to elevation and season is shown in Table~\ref{tab:elevation_summary}. Seasonal effects at each space are presented in Table~\ref{tab:seasonal_effects}.

\begin{table}[htbp]
	\centering
	\caption{Observed wind speed summary by elevation category and season.}
	\label{tab:elevation_summary}
	\begin{tabular}{llccc}
		\hline
		Elevation category & Season & Mean wind & SD & Number of observations \\
		\hline
		High ($>$1200 m)      & Autumn & 21.2 & 8.08 & 4140 \\
		Low ($<$300 m)        & Autumn & 20.3 & 7.62 & 2760 \\
		Medium (300--1200 m)  & Autumn & 19.7 & 6.82 & 1840 \\
		High ($>$1200 m)      & Spring & 24.2 & 7.66 & 4095 \\
		Low ($<$300 m)        & Spring & 23.5 & 7.83 & 2730 \\
		Medium (300--1200 m)  & Spring & 21.1 & 6.55 & 1820 \\
		High ($>$1200 m)      & Summer & 21.2 & 6.06 & 3987 \\
		Low ($<$300 m)        & Summer & 23.1 & 7.53 & 2658 \\
		Medium (300--1200 m)  & Summer & 19.4 & 5.17 & 1772 \\
		High ($>$1200 m)      & Winter & 26.5 & 10.9 & 4140 \\
		Low ($<$300 m)        & Winter & 22.1 & 9.28 & 2760 \\
		Medium (300--1200 m)  & Winter & 22.8 & 8.97 & 1840 \\
		\hline
	\end{tabular}
\end{table}

\begin{table}[htbp]
	\centering
	\caption{Estimated seasonal effects by spatial cluster.}
	\label{tab:seasonal_effects}
	\begin{tabular}{lccccccc}
		\hline
		Cluster & Autumn & Spring & Summer & Winter & Winter effect & Spring effect & Autumn effect \\
		\hline
		Coastal      & 20.33 & 23.52 & 23.11 & 22.12 & -0.99 & 0.42 & -2.78 \\
		Mid-altitude & 21.33 & 23.02 & 21.36 & 24.80 & 3.44 & 1.66 & -0.03 \\
		Interior     & 20.02 & 23.48 & 19.85 & 25.98 & 6.13 & 3.64 & 0.18 \\
		\hline
	\end{tabular}
\end{table}

Table~\ref{tab:seasonal_effects} presents the model-based seasonal estimates for the three spatial clusters. Coastal stations exhibited relatively small seasonal variation, with spring having the highest estimated wind speed (23.52 m/s), while autumn recorded the lowest. Mid-altitude stations experienced stronger winter conditions, with an estimated mean wind speed of 24.80 m/s, corresponding to a positive winter effect of 3.44 relative to the reference season. Interior stations showed the strongest seasonal contrast, with winter having the highest estimated wind speed (25.98 m/s) and the largest positive winter effect (6.13). These results indicate that seasonal influences on wind speed become more pronounced inland, whereas coastal regions experience comparatively moderate seasonal fluctuations, likely due to the moderating influence of the ocean.

In general, the descriptive statistics and model-based analysis indicate that both elevation and spatial location affect seasonal wind speed. Winter tends to have stronger winds, especially in high-elevation and interior areas.

\subsection{Temporal Residual model using Autoregressive Model (AR(1))}

To address temporal correlation in the residual series, an autoregressive model of order 1 (AR(1)) was applied to each station's residuals. The motivation behind this is the presence of empirical temporal correlation in the residual series and the need to keep the model simple and computationally feasible. The stationarity analysis indicated that the residual processes were broadly stationary, with the ADF test rejecting the unit-root null hypothesis for 18 of the 19 stations

This specification was further strengthened by the results in figure\ref{fig:diagnostic_plot} in the appendix. The ACF and PACF plots in figure\ref{fig:diagnostic_plot},  show no significant spikes, which proves that all serial correlation has been successfully removed. Additionally, the residuals fluctuate evenly around zero over time, confirming a constant variance and stationarity. Thirdly, the points on the Q-Q plot lie close to the diagonal, indicating that the errors follow a normal distribution.
In addition, \citep{gong2024ar1} established that autoregressive order-1 (AR(1)) processes accurately capture the intrinsic limits of predictability of wind speed time series, whereby the temporal correlation structure determines the forecast horizons. Recent studies have indicated \citep{lydia2025wind} that lower-order autoregressive models are appropriate for hourly and daily wind speed predictions, and autoregressive order one (AR(1)) or autoregressive order two (AR(2)) processes are optimal.

\subsubsection{AR(1) results for Temporal Residual model}
The estimates of the autoregressive coefficient ($\phi$) for each station are given in Table~\ref{tab:ar_coefficients}. The estimates of the autoregressive coefficients ranged from 0.3358 (Winburg) to 0.6411 (Rhodes), with a global mean of 0.4602. Given that all the estimated coefficients were positive and below one, there was a moderate positive temporal correlation in the residual series that satisfied the stationarity assumption for an AR(1) process.
\begin{table}[htbp]
	\centering
	\caption{AR(1) coefficients by station.}
	\begin{tabular}{lrrrr}
		\toprule
		\textbf{Station} & \textbf{$\phi$} & \textbf{Station} & \textbf{$\phi$} \\
		\midrule
		Alexander Bay & 0.479 & Napier & 0.455 \\
		Beaufort West & 0.527 & Noupoort & 0.533 \\
		Butterworth & 0.443 & Pofadder & 0.486 \\
		Calvinia & 0.521 & Rhodes & 0.641 \\
		Eston & 0.349 & Strydenburg & 0.370 \\
		Humansdorp & 0.366 & Sutherland & 0.598 \\
		Jozini & 0.540 & Upington & 0.431 \\
		Kuruman & 0.446 & Vredenburg & 0.401 \\
		Memel & 0.412 & Vredendal & 0.413 \\
		Winburg & 0.336 & \textbf{Mean} & \textbf{0.460} \\
		\bottomrule
	\end{tabular}
	\label{tab:ar_coefficients}
\end{table}
Using the global estimate ($\hat{\phi} = 0.4602$), the innovations for the temporal components of each station were calculated. Innovations are the unexplained part of the residual series. The average innovations of each station are summarised in Table~\ref{tab:innovations}.

\begin{table}[htbp]
	\centering
	\caption{Mean temporal innovations by station.}
	\label{tab:innovations}
	\begin{tabular}{lrrrr}
		\toprule
		\textbf{Station} & Mean innovation & \textbf{Station} & Mean innovation \\
		\midrule
		Alexander Bay & -0.2176 &Napier & 0.0769 \\
		Beaufort West & 0.1207& Noupoort & -0.5259 \\
		Butterworth & 0.2699 &Pofadder & 0.5610 \\
		Calvinia & -0.6744 &Rhodes & -1.1677 \\
		Eston & -0.9964 &Strydenburg & 0.9694 \\
		Humansdorp & -0.5848 &Sutherland & -0.6960 \\
		Jozini & -0.2334 &Upington & -0.2516 \\
		Kuruman & 0.5704 &Vredenburg & 0.4110 \\
		Memel & 1.4826 &Vredendal & 0.5258 \\
		Winburg & 0.3688 \\
		\hline
	\end{tabular}
\end{table}
\FloatBarrier
The mean innovations specific to each station were typically centred at zero; however, at some stations the mean innovations could be considerably positive or negative, representing local deviations from the temporal behaviour on average. The innovations served as the response variable in the Bayesian NNGP model.

\subsection{Bayesian NNGP Spatial Modelling}
\subsubsection{MCMC implementation}
The Bayesian latent NNGP model was estimated using three independent Markov chain Monte Carlo (MCMC) chains. Each chain consisted of 30\,000 iterations, with the first 15\,000 iterations discarded as burn-in. No thinning was applied (thinning interval = 1), allowing all retained posterior samples to be used for inference.

Convergence of the three chains was assessed using the Gelman-Rubin potential scale reduction factor (PSRF). The results are presented in Table~\ref{tab:gelman}.

\begin{table}[htbp]
	\centering
	\caption{Gelman-Rubin convergence diagnostics for the Bayesian NNGP model.}
	\label{tab:gelman}
	\begin{tabular}{lcc}
		\hline
		Parameter & PSRF & Upper 95\% CI \\
		\hline
		$\sigma^{2}$ & 1.00 & 1.00 \\
		$\tau^{2}$ & 1.00 & 1.00 \\
		$\phi$ & 1.00 & 1.00 \\
		$\nu$ & 1.01 & 1.02 \\
		\hline
		Multivariate PSRF & \multicolumn{2}{c}{1.01} \\
		\hline
	\end{tabular}
\end{table}

All PSRF values were very close to one, with the largest value being 1.01 for the smoothness parameter ($\nu$).Since all the values were well below the convergence criterion of 1.10, the MCMC chains converged successfully, indicating that the posterior distributions were reliable and independent of any initial values assigned.

The sampling efficiency was tested through the effective sample size (ESS).

\begin{table}[htbp]
	\centering
	\caption{Effective sample sizes for the Bayesian NNGP parameters.}
	\label{tab:ess}
	\begin{tabular}{lc}
		\hline
		Parameter & Effective sample size \\
		\hline
		$\sigma^{2}$ & 15\,319.5 \\
		$\tau^{2}$ & 10\,168.3 \\
		$\phi$ & 1\,988.1 \\
		$\nu$ & 810.1 \\
		\hline
	\end{tabular}
\end{table}

The effective sample size indicates that the MCMC procedure generated many independent draws from the posterior for all model parameters. The sampling efficiency of variance parameters ($\sigma^{2}$ and $\tau^{2}$) was especially high, whereas the sampling efficiency of the spatial decay ($\phi$) and smoothness ($\nu$) parameters was relatively low. Figures \ref{fig:density_mcmc} and \ref{fig:trace_mcmc} illustrate MCMC diagnostics for the spatial parameters in the Bayesian NNGP model. The trace and density plots illustrate very good chain mixing and cross-chain convergence for the parameters, corroborating the high-quality ESS statistics provided in Table~\ref{tab:ess}. 
The overall convergence diagnostics and effective sample size results imply that the MCMC sampling method produced stable and efficient parameter estimates of the Bayesian NNGP model.

\subsubsection{Posterior parameter estimates}
The NNGP spatial model was fit to the AR(1) residuals at all stations. The posterior estimates of the spatial parameters are presented in Table~\ref{tab:spatial_params}.

\begin{table}[htbp]
	\centering
	\caption{Posterior parameter estimates.}
	\begin{tabular}{lrrrr}
		\toprule
		\textbf{Parameter} & \textbf{Mean} & \textbf{SD} & \textbf{2.5\%} & \textbf{97.5\%} \\
		\midrule
		$\sigma^2$ (partial sill) & 0.391 & 0.171 & 0.160 & 0.816 \\
		$\tau^2$ (nugget) & 0.246 & 0.134 & 0.082 & 0.589 \\
		$\phi$ (decay) & 5.260 & 2.731 & 0.673 & 9.743 \\
		$\nu$ (smoothness) & 1.251 & 0.422 & 0.543 & 1.960 \\
		Range (km) & 102.878 & 100.339 & 34.185 & 423.785 \\
		\bottomrule
	\end{tabular}
	\label{tab:spatial_params}
\end{table}

The range parameter indicates that spatial correlation decays with distance, with the effective range approximately 103 km. The smoothness parameter $\nu \approx 1.25$ suggests moderate smoothness of the process, intermediate between exponential ($\nu = 0.5$) and squared-exponential ($\nu \to \infty$).

\subsection{Comparison of Forecasting Models}

Several forecasting models were considered to identify which model would be appropriate for wind speed predictions. Some of the forecasting models considered include regression models, the AR (1) model and hybrid models which include time and space. The accuracy of the models was based on the RMSE (Root Mean Squared Error).

\begin{table}[htbp]
	\centering
	\caption{Comparison of forecasting models based on RMSE.}
	\label{tab:model_comparison}
	\begin{tabular}{clc}
		\hline
		Rank & Model & RMSE \\
		\hline
		1 & AR(1)-NNGP residual correction & 1.416 \\
		2 & AR(1) & 1.715 \\
		3 & AR(1)-NNGP + elevation + cluster & 1.559 \\
		4 & Regression + NNGP & 2.147 \\
		5 & Regression & 2.442 \\
		\hline
	\end{tabular}
\end{table}

The results in Table~\ref{tab:model_comparison} demonstrate that the AR(1)-NNGP residual correction model achieved the lowest RMSE (1.416), indicating the highest forecasting accuracy among the evaluated approaches. The standalone ARIMA model achieved RMSE = 1.715, confirming the importance of modelling temporal dependence in wind speed observations.

The addition of spatial corrections through the NNGP resulted in a substantial improvement over AR(1) alone, reducing the RMSE by approximately 17.4\%. This indicates that the residual spatial structure not captured by the temporal model contains valuable predictive information.

Interestingly, the AR(1)-NNGP model including elevation and cluster covariates produced a slightly higher RMSE (1.559) compared with the residual correction approach. This suggests that explicitly including these covariates did not provide additional predictive benefit beyond the spatial residual process captured by the NNGP.

The regression-based approaches performed comparatively worse, with RMSE values of 2.147 and 2.442 for Regression + NNGP and Regression alone, respectively. These results highlight the importance of accounting for temporal dependence when forecasting wind speed, as regression models alone are unable to adequately represent the dynamic behaviour of the wind process.

Based on these results, the AR(1)-NNGP residual correction model was selected as the final forecasting model and was subjected to further evaluation using horizon-specific errors, station-level performance, statistical forecast comparison tests, and conformal prediction intervals.

\subsection{Evaluation of the Hybrid Model}

\paragraph{Station-Specific Spatial Corrections from the NNGP Model}

Following the identification of the AR(1)-NNGP residual correction model as the best-performing forecasting approach, the estimated spatial corrections were examined to determine whether the NNGP component was capturing meaningful spatial variability. The results for each station are presented in Table~\ref{tab:spatial_corrections}.
\begin{table}[htbp]
	\centering
	\caption{Station-specific spatial corrections estimated by the NNGP model.}
	\label{tab:spatial_corrections}
	\begin{tabular}{lrrrr}
		\toprule
		\textbf{Station} & Spatial correction & \textbf{Station} & Spatial correction \\
		\midrule
		Alexander Bay & -0.2090 &Napier & -0.2130 \\
		Beaufort West & -0.5310& Noupoort & 0.8170 \\
		Butterworth & -0.8400 &Pofadder & -0.2730 \\
		Calvinia & -0.6680 &Rhodes & 1.7270 \\
		Eston & -0.4880 &Strydenburg & -0.080 \\
		Humansdorp & -0.4540 &Sutherland & -0.3180 \\
		Jozini & 0.5400 &Upington & -0.5480 \\
		Kuruman & -0.2830 &Vredenburg & -0.1220 \\
		Memel & 0.8340 &Vredendal & 0.0900 \\
		Winburg & 0.9950 \\
		\hline
	\end{tabular}
\end{table}

The estimated spatial corrections varied substantially among stations, ranging from $-0.840$ at Butterworth to $1.727$ at Rhodes. Positive corrections indicate locations where the AR(1) model systematically underestimated wind speed, while negative corrections indicate locations where AR(1) forecasts tended to overestimate wind speed. Therefore, the NNGP component adjusts the forecasts according to the local spatial characteristics of each station.

The presence of both positive and negative corrections demonstrates that the NNGP model is not simply applying a constant bias adjustment. Instead, it captures spatially varying residual patterns associated with station location. This confirms that the improvement achieved by the hybrid model results from modelling residual spatial dependence that remains after accounting for temporal dynamics through AR(1).
\begin{figure}[H]
	\centering
	\includegraphics[width=0.8\textwidth]{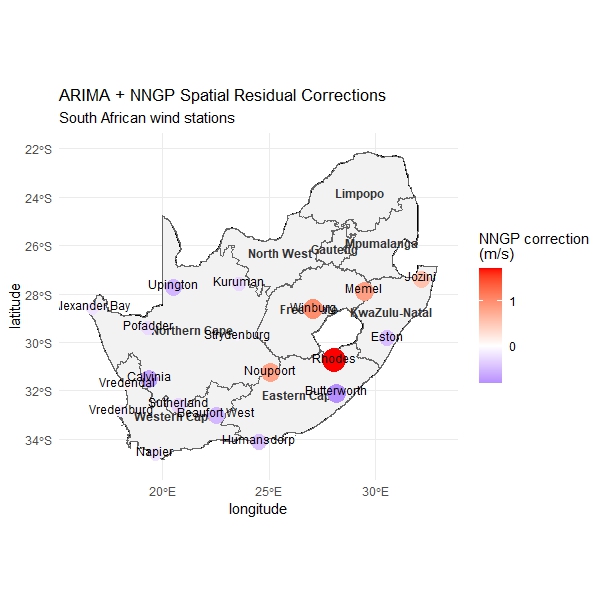}
	\caption{Posterior mean NNGP spatial residual corrections for the 19 meteorological stations.}
	\label{fig:spatial_correction}
\end{figure}

The spatial residual correction for the South African wind monitoring network is shown in Figure \ref{fig:spatial_correction}. A positive spatial correction (red) is seen in the location where the AR(1) model underestimated the wind speed. On the other hand, a negative spatial correction (blue) is observed at locations where the AR(1) model overestimated the wind speed. The magnitude of each point corresponds to the value of the spatial correction added to the AR(1) prediction.

The geographical patterns in the spatial correction surface show that there are no random errors in the spatial correction. Rhodes shows the largest positive correction, along with positive corrections in Winburg, Memel, Noupoort, and Jozini. The stations mentioned above are mostly located in the high-elevation areas of South Africa. This implies that the AR(1) model failed to capture the topographic effect and regional atmospheric circulation on wind speed.

\paragraph{Station-Specific Performance}
Station-specific improvements varied considerably (Table~\ref{tab:station_rmse}) and Figure \ref{fig:station_rmse} in the appendix, with the largest gains observed at high-elevation stations.

\begin{table}[htbp]
	\centering
	\caption{Station-level RMSE and improvement}
	\begin{tabular}{lrrr}
		\toprule
		\textbf{Station} & \textbf{ARIMA} & \textbf{Hybrid} & \textbf{Improvement (\%)} \\
		\midrule
		Alexander Bay & 1.992 & 1.916 & 3.8 \\
		Beaufort West & 1.799 & 1.446 & 19.6 \\
		Butterworth & 2.635 & 2.134 & 19.0 \\
		Calvinia & 1.713 & 1.134 & 33.8 \\
		Eston & 1.777 & 1.473 & 17.1 \\
		Humansdorp & 1.639 & 1.336 & 18.5 \\
		Jozini & 1.258 & 1.222 & 2.8 \\
		Kuruman & 1.726 & 1.586 & 8.1 \\
		Memel & 1.620 & 1.392 & 14.1 \\
		Napier & 1.287 & 1.161 & 9.8 \\
		Noupoort & 1.435 & 1.199 & 16.4 \\
		Pofadder & 1.429 & 1.271 & 11.0 \\
		\textbf{Rhodes} & \textbf{2.536} & \textbf{1.297} & \textbf{48.9} \\
		Strydenburg & 1.234 & 1.195 & 3.1 \\
		Sutherland & 1.307 & 1.083 & 17.1 \\
		Upington & 1.900 & 1.555 & 18.2 \\
		Vredenburg & 1.429 & 1.374 & 3.8 \\
		Vredendal & 1.473 & 1.500 & -1.8 \\
		Winburg & 1.590 & 1.177 & 26.0 \\
		\bottomrule
	\end{tabular}
	\label{tab:station_rmse}
\end{table}

The substantial improvement at Rhodes (48.9\%) reflects the strong spatial signal at this high-altitude station. The minimal improvement at Vredendal (-1.8\%) suggests that the spatial correction was not beneficial for this coastal station, possibly due to weaker spatial coherence with other stations.

\paragraph{Horizon-Wise Performance}

\begin{figure}[htbp]
	\centering
	\includegraphics[width=0.8\textwidth]{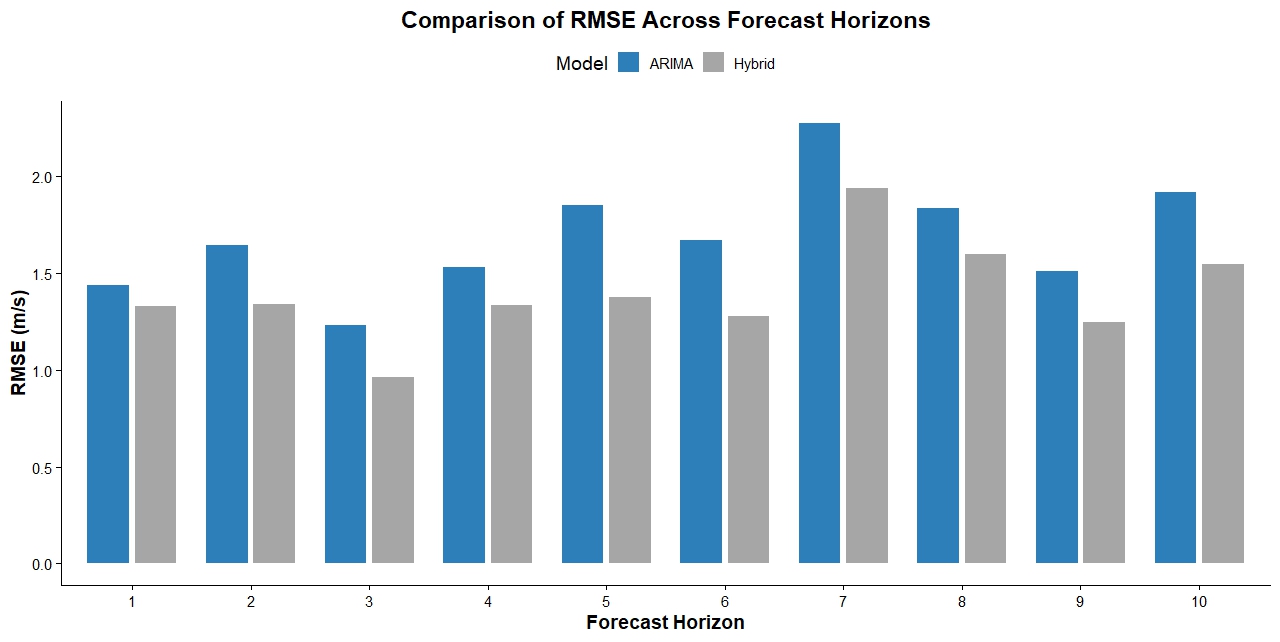}
	\caption{Horizon-wise RMSE (m/s).}
	\label{fig:horizon_rmse}
\end{figure}

The performance of both models varied with forecast horizon (Table~\ref{tab:horizon_rmse}) also presented in (Figure \ref{fig:horizon_rmse}) . The improvement is relatively stable across horizons, with no evidence of decay in the spatial correction's effectiveness. This result is consistent with the decay analysis, which found that the constant spatial correction ($\lambda = 0$) was optimal.

\begin{table}[htbp]
	\centering
	\caption{Horizon-wise RMSE (m/s).}
	\begin{tabular}{lrrr}
		\toprule
		\textbf{Horizon} & \textbf{ARIMA} & \textbf{Hybrid} & \textbf{Improvement (\%)} \\
		\midrule
		1 & 1.440 & 1.328 & 7.8 \\
		2 & 1.645 & 1.338 & 18.6 \\
		3 & 1.234 & 0.963 & 22.0 \\
		4 & 1.532 & 1.336 & 12.8 \\
		5 & 1.853 & 1.375 & 25.8 \\
		6 & 1.671 & 1.280 & 23.4 \\
		7 & 2.275 & 1.942 & 14.6 \\
		8 & 1.839 & 1.597 & 13.2 \\
		9 & 1.510 & 1.245 & 17.6 \\
		10 & 1.922 & 1.547 & 19.5 \\
		\bottomrule
	\end{tabular}
	\label{tab:horizon_rmse}
\end{table}

\paragraph{Conformal Prediction Intervals}
Blockwise conformal prediction was used to quantify uncertainty and the results are summarised in Table~\ref{tab:conformal}. A nominal level of $\alpha=0.10$ was used, corresponding to a target coverage of $90\%$. A block size of $B=7$ produced lower empirical coverage while remaining above the nominal $90\%$ target. This provided a more favourable balance between coverage and interval sharpness. 
The 7-day block produced approximately $99.5\%$ coverage against a nominal 90\% target, while giving an average interval width of $7.018$~m/s. 

\begin{table}[htbp]
	\centering
	\caption{Conformal prediction performance.}
	\begin{tabular}{lr}
		\toprule
		\textbf{Metric} & \textbf{Value} \\
		\midrule
		Best Block size & 7 \\
		Calibration residual quantile (q90) & 3.509 m/s \\
		Nominal coverage & 90\% \\
		Achieved coverage & 99.5\% \\
		Average interval width & 7.018 m/s \\
		\bottomrule
	\end{tabular}
	\label{tab:conformal}
\end{table}

\begin{figure}[htbp]
	\centering
	\includegraphics[width=0.8\textwidth]{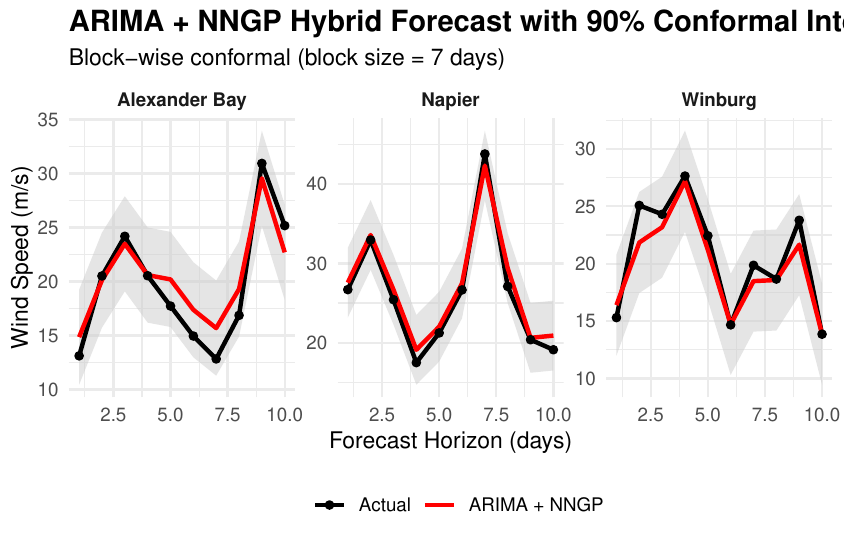}
	\caption{Conformal prediction plot for selected stations.}
	\label{fig:conformal_plot}
\end{figure}
The predictive capability of the hybrid model was evaluated using block-wise conformal prediction intervals (figure~\ref{fig:conformal_plot} and figure~\ref{fig:conformal_all}). 
The prediction intervals maintain a consistent width across the 10-day forecast window. This proves that the model stops errors from compounding as it predicts further out. The hybrid forecast in figure~\ref{fig:conformal_plot}, (red line) tracks the actual wind speed data points (black dots) with high precision. It successfully captures sudden regional wind surges, such as the peak at Napier around Day 7. The NNGP model's spatial adjustments successfully correct localized biases, allowing the forecast to accurately match observations at stations with high spatial variance like Winburg. Although the 99.5\% achieved coverage is statistically conservative, it provides absolute security for wind farm grid scheduling by ensuring extreme wind events almost never breach the predicted safety bands. 

\paragraph{Diebold-Mariano Test Results}
Table~\ref{tab:dm_results} and Figure \ref{fig:dm_plot} presents the results of the Diebold-Mariano (DM) test comparing the forecasting accuracy of the ARIMA baseline model with that of the proposed hybrid model across the ten forecast horizons. The DM test evaluates whether the difference in forecast errors between two competing models is statistically significant, with a positive and statistically significant DM statistic indicating superior predictive performance of the hybrid model.

\begin{table}[htbp]
	\centering
	\caption{Diebold-Mariano test by horizon.}
	\begin{tabular}{lrrr}
		\toprule
		\textbf{Horizon} & \textbf{DM Statistic} & \textbf{p-value} & \textbf{Significant} \\
		\midrule
		1 & 0.985 & 0.337 & No \\
		2 & 2.116 & 0.048 & Yes \\
		3 & 1.915 & 0.071 & No \\
		4 & 1.534 & 0.142 & No \\
		5 & 3.398 & 0.003 & Yes \\
		6 & 2.060 & 0.054 & No \\
		7 & 2.356 & 0.030 & Yes \\
		8 & 1.724 & 0.102 & No \\
		9 & 1.905 & 0.073 & No \\
		10 & 2.180 & 0.043 & Yes \\
		\bottomrule
	\end{tabular}
	\label{tab:dm_results}
\end{table}

The pooled DM test across all horizons was highly significant:
\begin{itemize}
	\item \textbf{DM statistic}: 6.44
	\item \textbf{p-value}: $<$ 0.001
\end{itemize}
Unlike the horizon-specific tests, which evaluate performance independently at each forecast lead time, the pooled analysis assesses the overall predictive ability of the competing models across the complete forecasting horizon. The highly significant pooled result confirms that the observed reductions in forecasting error are systematic rather than occurring by chance. This confirms that the hybrid model provides statistically significant improvements over the AR(1) baseline.

\paragraph{Overall model performance}
Summary of the AR(1)-NNGP residual correction model performance and parameter estimates.
\begin{table}[htbp]
	\centering
	\caption{Summary of the AR(1)-NNGP residual correction model.}
	\label{tab:arima_nngp_summary}
	\begin{tabular}{lll}
		\hline
		\textbf{Component} & \textbf{Metric} & \textbf{Value} \\
		\hline
		
		\textbf{Temporal Component} 
		& Model structure & ARIMA residual modelling \\
		
		& Global AR(1) coefficient ($\phi$) 
		& 0.4602 \\
		
		& Station-specific AR(1) range 
		& 0.3358--0.6411 \\
		
		\hline
		
		\textbf{Spatial NNGP Component} 
		& Covariance model 
		& Mat\'ern NNGP \\
		
		& Number of neighbours 
		& 8 \\
		
		& Partial sill ($\sigma^2$) 
		& 0.3903 \\
		
		& Nugget variance ($\tau^2$) 
		& 0.2451 \\
		
		& Decay parameter ($\phi$) 
		& 5.2618 \\
		
		& Smoothness parameter ($\nu$) 
		& 1.2389 \\
		
		& Effective spatial range 
		& 102.9 km \\
		
		& Spatial residual variance proportion 
		& 61.4\% \\
		
		\hline
		
		\textbf{Forecast Performance} 
		& ARIMA + NNGP RMSE 
		& 1.416 m/s \\
		
		& ARIMA RMSE 
		& 1.715 m/s \\
		
		& Improvement over ARIMA 
		& 17.4\% \\
		
		\hline
		
		\textbf{Statistical Validation} 
		& Diebold--Mariano statistic 
		& 6.4804 \\
		
		& Significance 
		& $p<0.001$ \\
		
		\hline
		
		\textbf{Uncertainty Quantification} 
		& Method 
		& Blockwise Conformal prediction \\
		
		& Empirical coverage 
		& 99.95\% \\
		
		& Mean interval width 
		& 7.018 m/s \\
		
		\hline
	\end{tabular}
\end{table}

\section{Discussion} \label{sec:5}
\subsection{Summary of Key Findings}

This study developed a hybrid spatio-temporal forecasting framework that combines autoregressive integrated moving average of order 1 (AR(1)) models with Bayesian nearest-neighbour Gaussian process (NNGP) residual correction for wind speed forecasting. 
The results demonstrate that wind speed variability contains both temporal persistence and spatial dependence. The AR(1) component successfully captured temporal behaviour, while the NNGP model identified and corrected residual spatial patterns that remained after temporal modelling. The integration of these two components resulted in improved forecast accuracy, statistically significant improvements over AR(1) alone, and reliable uncertainty quantification through conformal prediction intervals.

\subsection{Summary of Major Findings}

\begin{itemize}
	\item The AR(1)-NNGP residual correction model achieved the best forecasting performance with an overall RMSE of 1.416, outperforming all alternative models.
	
	\item AR(1) alone improved upon regression-based approaches, confirming that wind speed contains significant temporal dependence.
	
	\item Bayesian NNGP modelling identified substantial residual spatial dependence after AR(1) correction, with an estimated effective spatial range of approximately 
	$102.88 km$.
	
	\item Approximately 61\% of residual variability was attributed to spatially structured variation, demonstrating that AR(1) errors contain exploitable spatial information rather than random noise alone.
	
	\item The hybrid model significantly outperformed AR(1) according to the 
	Diebold--Mariano test ($DM=6.48$, $p<0.001$).
	
	\item Blockwise Conformal prediction intervals achieved approximately 99.5\% empirical coverage, demonstrating reliable uncertainty quantification. A 7-day block aligns with the natural weekly cycle of wind patterns. This alignment lets the algorithm model the time-series dependencies accurately, resulting in a slightly tighter safety band that leaves a 0.5\% margin of error.
	
	\item Seasonal analysis indicated stronger wind conditions during winter and spring, particularly in high-elevation and interior regions, suggesting potential periods 
	where maintenance activities may be more challenging.
\end{itemize}

\subsection{Overall Discussion}

Initially, the analysis involved setting up benchmarks for temporal forecasts through regression and AR(1) models. While regression did well in capturing the general relationships between wind speeds and the environmental predictors, the model was limited in its ability to forecast due to poor capture of temporal persistence. AR(1) addressed this problem by capturing dependence in wind speed anomalies, with autoregressive coefficients estimated to be positive for all stations and a global coefficient of about 0.46. This means that historical wind speeds can predict future states. However, the persistence in the residuals obtained from the AR(1) model indicates that there is a need for spatial correction.

To address this, a Bayesian nearest-neighbour Gaussian process (NNGP) was developed to correct the ARIMA residuals. Posterior results showed spatial dependence, with a partial sill of 0.390, a nugget variance of 0.250, and an effective spatial range of about 103 km, suggesting that the wind pattern residuals are correlated at the regional scale. The spatially structured variance proportion was 0.609, indicating that approximately 61\% of the variance is explained by spatial structure, with the remaining 39\% attributable to local effects. This proves that the conclusion about the improvements resulting from spatial learning in the hybrid model is valid. In addition, station-specific adjustments varied, suggesting that the NNGP model accounts for local spatial heterogeneity.

The hybrid model was then tested against other models using stringent testing methods. It performed the best among all others in terms of having the least root mean square error for all forecast intervals and locations, even compared to the individual AR(1) model. Significance was established using the Diebold-Mariano tests, with a pooled statistic of 6.48 and an associated p-value of less than 0.001, indicating that the hybrid model is significantly better at forecasting. This shows the importance of breaking down the problem of forecasting into two problems – the temporal and the spatial problem.

Regarding uncertainty quantification, it was performed using conformal prediction intervals, which yielded a coverage rate of around 99.5\%, providing sufficient evidence of the model's calibration. The interval length was 7.018 m/s, providing a good balance between uncertainty and forecast accuracy. Interval predictions are needed in practice because point forecasts might conceal some risks.

The analysis of seasonality yields insights that can help manage wind energy operations. The strongest winds are expected during the winter and spring months in inland and elevated areas. Even though it might facilitate energy production, it also increases maintenance needs. On the contrary, the summer months are usually characterised by weaker winds and therefore might provide better opportunities to carry out planned maintenance operations. However, other factors need to be considered, such as electricity demand, network limitations, precipitation, and the weather forecasts for the next few days. Finally, the estimated correlation range of about 103 km might be useful for network design monitoring.

\section{Conclusion} \label{sec:6}

This study developed a hybrid AR(1)-NNGP forecasting framework for wind speed prediction by combining temporal modelling with Bayesian spatial residual correction. The study demonstrated that wind variability cannot be fully described by temporal models alone because residual spatial dependence remains after AR(1) forecasting.

The first objective was achieved through the development of regression and AR(1) baselines. The results showed that temporal persistence plays an important role in wind forecasting, with AR(1) outperforming regression approaches.

The second objective was achieved through Bayesian NNGP residual modelling. The spatial model successfully identified structured residual variability, with approximately 61\% of residual variance explained by spatial dependence and an effective spatial correlation range of approximately 103 km.

The third and fourth goals were achieved through validation. The AR(1)-NNGP residual correction model proved to be the best-performing model, with an RMSE of 1.416, significantly outperforming the AR(1) model according to the Diebold-Mariano test.
The fifth goal was accomplished using conformal prediction to obtain calibrated uncertainty estimates for forecasts.

In general, the results clearly show that combining temporal and spatial modelling yields a better solution for wind speed forecasting. The presented framework not only provides better forecasting results but also calibrates forecast uncertainty, making it useful as a decision-support tool for wind energy.

\subsection{Recommendations for Future Work}

Taking into consideration the above-mentioned results, several directions for future research can be suggested:

\begin{enumerate}
	\item \textbf{Application at higher temporal resolution}: Application of the framework to the hourly wind speed data will allow us to investigate sub-daily patterns.
	\item \textbf{Integration with meteorological forecasts}: The use of numerical weather predictions as covariates will allow us to improve the baseline model.
	
	\item \textbf{Operational implementation}: Developing a real-time forecasting system based on the hybrid framework for operational use in wind energy management.
\end{enumerate}

Nevertheless, this study demonstrates that AR(1)-NNGP residual correction provides a powerful and interpretable framework for improving wind speed forecasts across spatially distributed monitoring networks.
%\section*{Acknowledgements}
%\section{Results}\label{sec2}

%=====================================================================
\section*{Declarations}
\begin{itemize}
\item \textbf{Funding.} This research received no specific grant from any funding agency in the public, commercial, or not-for-profit sectors.
\item \textbf{Conflict of interest.} The authors declare no conflicts of interest.
\item \textbf{Ethical approval.} This article does not contain any studies with human participants or animals performed by any of the authors.
\item \textbf{Data availability.} The data in brief can be accessed at \url{https://github.com/csigauke/AHybrid-Linear--Gaussian-Process-Framework-for-Spatio-Temporal-Wind-Speed-Forecasting} (accessed on 16 January 2025).
\item \textbf{Codes.} The code supporting the findings of this study is available upon request.
\item \textbf{Author contributions.} Conceptualisation, T.H.T.\ and C.S.; methodology, T.H.T.; software, T.H.T.; validation, T.H.T., C.S., T.B.D., and S.N.; formal analysis, T.H.T.; investigation, T.H.T.\ and C.S.; writing---original draft preparation, T.H.T.; writing---review and editing, T.H.T., C.S., T.B.D., and S.N.; supervision, C.S., T.B.D., and S.N. All authors have read and agreed to the published version of the manuscript.
\end{itemize}

%=====================================================================
\appendix
\section{Supplementary Figures}

\begin{figure}[H]
	\centering
	\includegraphics[width=0.8\textwidth]{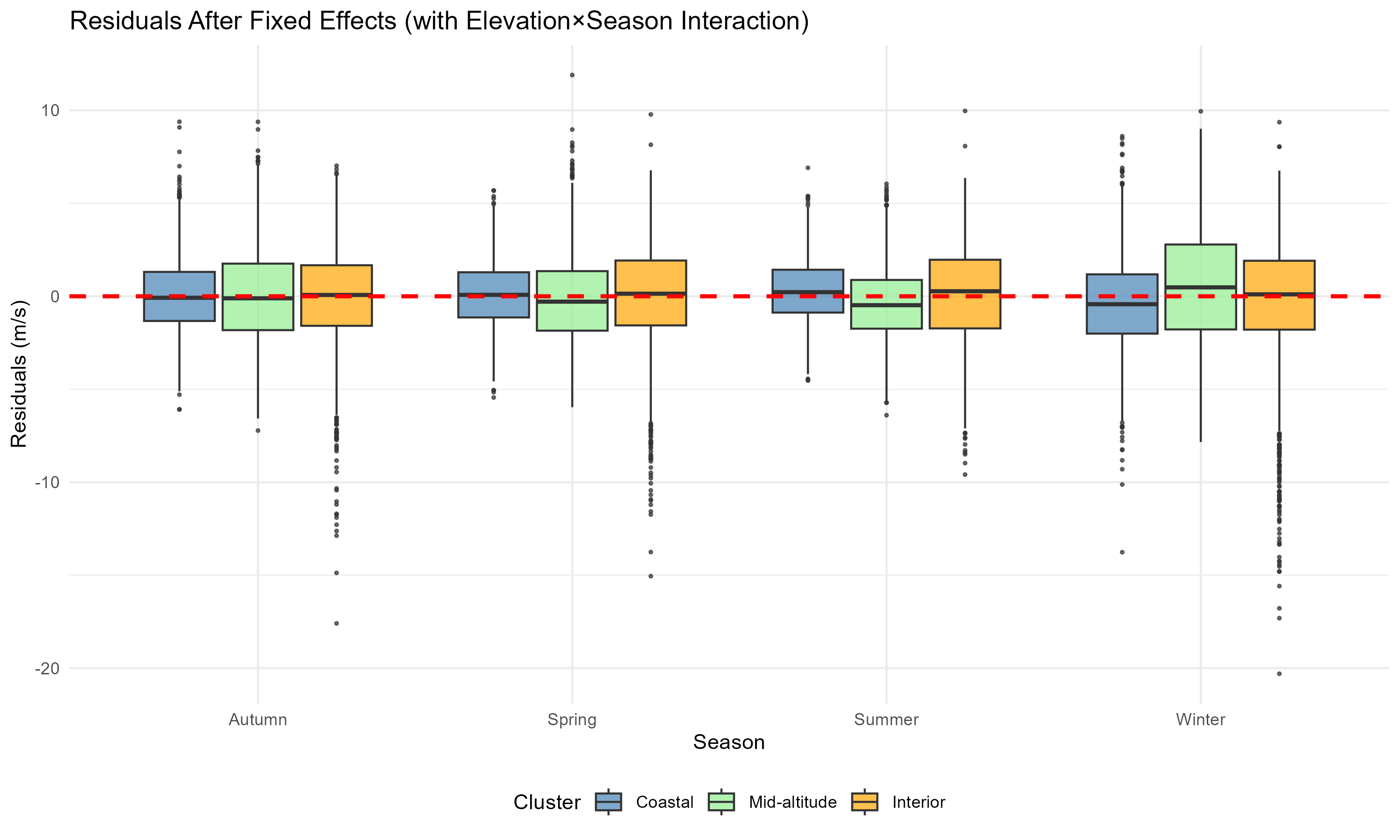}
	\caption{Seasonal residuals plot.}
	\label{fig:residual_wind}
\end{figure}

\begin{figure}[H]
	\centering
	\includegraphics[width=0.8\textwidth]{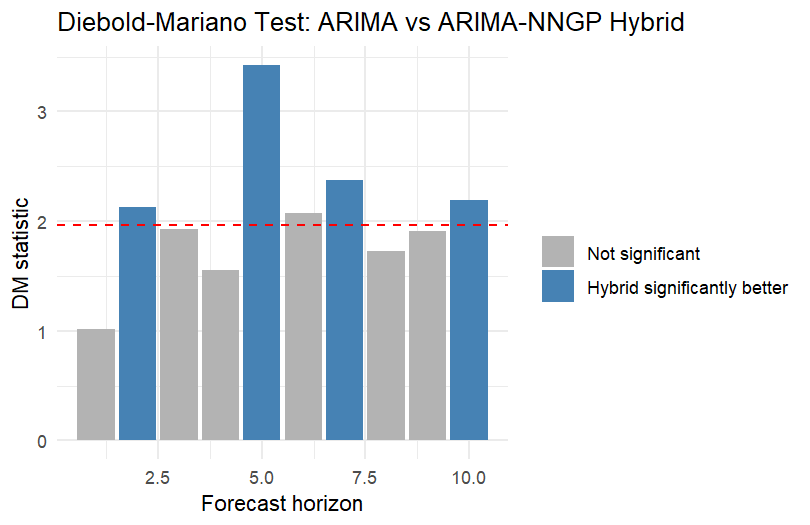}
	\caption{Diebold-Mariano horizon plot.}
	\label{fig:dm_plot}
\end{figure}

\begin{figure}[H]
	\centering
	\includegraphics[width=0.8\textwidth]{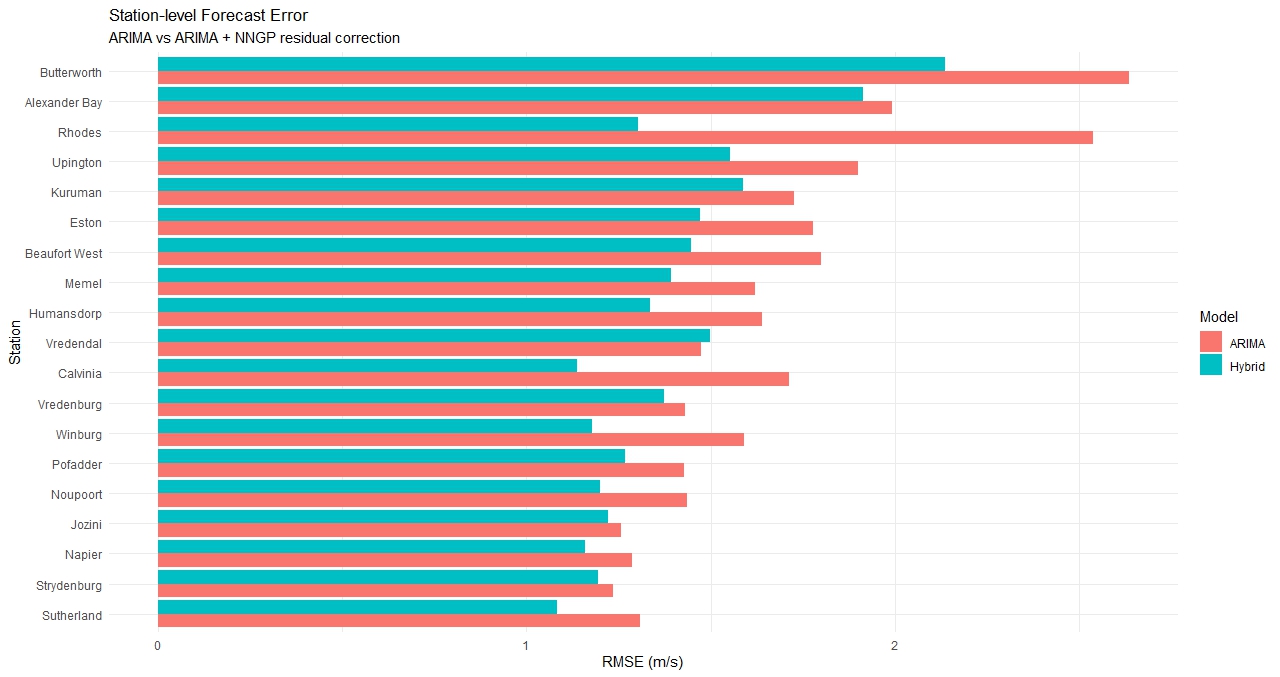}
	\caption{Stationwise RMSE.}
	\label{fig:station_rmse}
\end{figure}

\begin{figure}[H]
	\centering
	\includegraphics[width=0.8\textwidth]{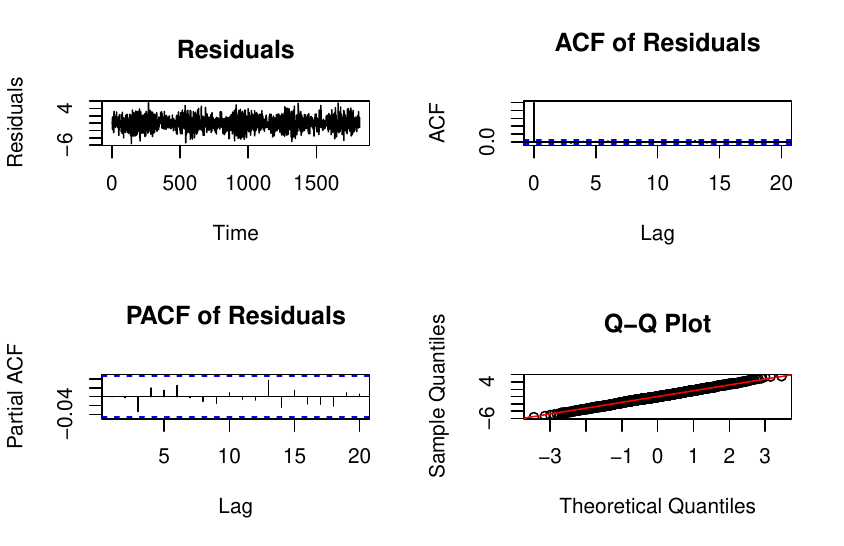}
	\caption{Diagnostics Plot.}
	\label{fig:diagnostic_plot}
\end{figure}

\begin{figure}[H]
	\centering
	\includegraphics[width=0.8\textwidth]{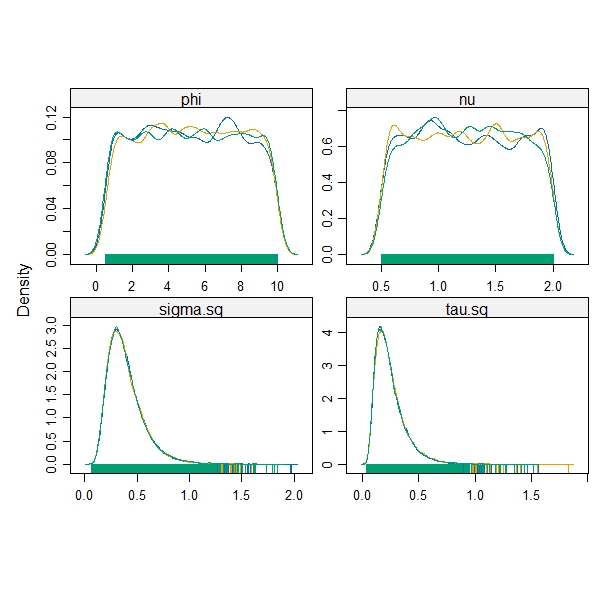}
	\caption{Density plots.}
	\label{fig:density_mcmc}
\end{figure}

\begin{figure}[H]
	\centering
	\includegraphics[width=0.8\textwidth]{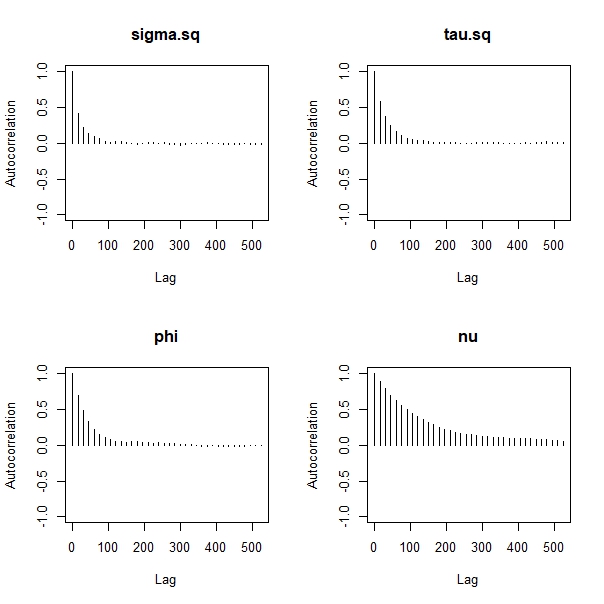}
	\caption{Trace plots.}
	\label{fig:trace_mcmc}
\end{figure}

\begin{figure}[H]
	\centering
	\includegraphics[width=0.8\textwidth]{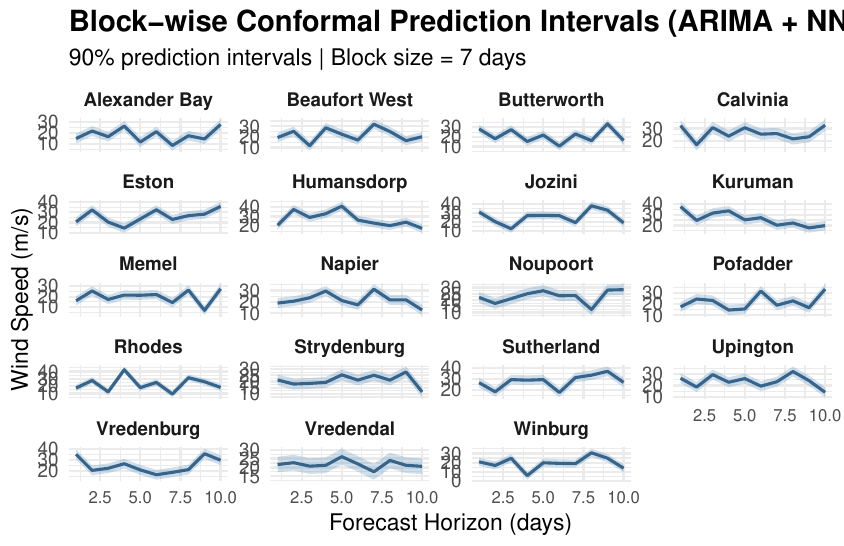}
	\caption{Conformal plot for all stations.}
	\label{fig:conformal_all}
\end{figure}

%\begin{figure}[H]
%\centering
%\includegraphics[width=0.8\textwidth]{density_mcmc.pdf}
%\caption{Posterior density plots for the Bayesian NNGP spatial parameters.}
%\label{fig:density_mcmc}
%\end{figure}
%
%\begin{figure}[H]
%\centering
%\includegraphics[width=0.8\textwidth]{trace_mcmc.pdf}
%\caption{Trace plots for the Bayesian NNGP spatial parameters.}
%\label{fig:trace_mcmc}
%\end{figure}
%
%\begin{figure}[H]
%\centering
%\includegraphics[width=0.8\textwidth]{station_rmse.pdf}
%\caption{Station-level RMSE comparison between ARIMA and the hybrid model.}
%\label{fig:station_rmse}
%\end{figure}
%
%\begin{figure}[H]
%\centering
%\includegraphics[width=0.8\textwidth]{conformal_all.pdf}
%\caption{Conformal prediction intervals for all stations.}
%\label{fig:conformal_all}
%\end{figure}
%
%\begin{figure}[H]
%\centering
%\includegraphics[width=0.8\textwidth]{dm_plot.pdf}
%\caption{Diebold--Mariano test statistics by forecast horizon.}
%\label{fig:dm_plot}
%\end{figure}

%%=====================================================================
%% References are generated by BibTeX from the attached sn-bibliography.bib.
%% Compile sequence: pdflatex -> bibtex -> pdflatex -> pdflatex
%%=====================================================================
%\bibliographystyle{plainnat}
%\bibliography{sn-bibliography}

\end{document}